\documentclass[seceq]{ptptex}
\usepackage{bm}
\usepackage[dvips]{graphicx}
\newcommand{\Slash}[1]{\ooalign{\hfil/\hfil\crcr$#1$}}

\newcommand{\im}{\text{Im}}
\newcommand{\Tr}{\text{Tr}}

\notypesetlogo                       

\title{
Detailed Analysis of the Chiral Unitary
Model for Meson-Baryon Scattering
with Flavor $SU(3)$ Breaking Effects}

\author{
Tetsuo \textsc{Hyodo}$^{1}$,
Seung-il \textsc{Nam}$^{1,2}$,
Daisuke \textsc{Jido}$^{1,}$\footnote{Present address:
ECT*, European Centre for Theoretical Studies
in Nuclear Physics and Related Areas
Villa Tambosi, Strada delle Tabarelle 286,
I-38050 Villazzano (Trento), Italy.} and
Atsushi \textsc{Hosaka}$^{1}$%
}

\inst{
$^{1}$Research Center for Nuclear Physics (RCNP), 
Ibaraki, 567-0047, Japan \\
$^{2}$Department of Physics,
Pusan National University, Pusan 609-735, Korea
}

\abst{
We study $s$-wave meson-baryon scattering
using the chiral unitary model. 
We consider $1/2^{-}$ baryon resonances
as quasibound states
of the low lying mesons ($\pi,K,\eta$)
and baryons ($N,\Lambda,\Sigma,\Xi$).
In previous works, the subtraction constants
which appear in loop integrals were found to largely depend on the
channels,
and it was necessary to fit these constants to reproduce the data.
In order to extend this model to all channels 
with fewer parameters,
we introduce flavor $SU(3)$ breaking interactions
in the framework of chiral perturbation theory.
It is found, however, that the observed $SU(3)$ breaking in
meson-baryon scattering cannot be explained 
by the present $SU(3)$ breaking interactions.
The role and importance of the subtraction constants
in the present framework are discussed.
}

\begin{document}

\maketitle

\section{Introduction}\label{sec:intro}

A unified 
study of meson-baryon scattering
in various channels 
is important to understand hadron dynamics
in low and intermediate energy regions
from the viewpoint of QCD.
In particular, 
the properties of excited states of baryons
observed in meson-baryon scattering as resonances
have been investigated with great interest 
both theoretically and experimentally.
At this time, there are several established approaches to describe
the properties of baryon resonances.
A recent development in this field is the introduction of the
chiral unitary model~\cite{Kaiser:1995cy,Kaiser:1995eg,Kaiser:1997js,
Krippa:1998us,Oset:1998it,Lutz:2001yb}, in which
the $s$-wave baryon resonances are dynamically generated
in meson-baryon scattering,
while the conventional quark model approach
describes the baryon resonances as three-quark
states with an excitation of one of the quarks.

The chiral unitary model is based on 
chiral perturbation theory (ChPT)~\cite{Weinberg:1979kz,
Gasser:1985gg}.
Imposing the unitarity condition,
we can apply the ChPT in regions of higher energy 
than in the original perturbative calculation,
and we can study properties of resonances generated by
non-perturbative resummations.
In the implementation of the unitarity condition,
regularization of loop integrals 
brings parameters into this model, such as the three-momentum cut-off
and the ``subtraction constants'' in the dimensional 
regularization.

In Refs.~\citen{Kaiser:1995cy} and \citen{Oset:1998it},
$s$-wave scattering in meson
and baryon systems with strangeness $S=-1$ was investigated
by solving the Lippman-Schwinger equation in coupled channels,
where the $\Lambda(1405)$ resonance is dynamically generated
by meson-baryon scattering. 
In the regularization procedure, 
a form factor associated with the extended structure of hadrons
is introduced in the kernel potential obtained by
ChPT~\cite{Kaiser:1995cy},
while the loop integral is cut off
in the three-momentum~\cite{Oset:1998it}.
In Refs.~\citen{Oset:2001cn,Inoue:2001ip,Ramos:2002xh},
they extended the chiral unitary approach 
to other strangeness channels and obtained the baryonic resonances
$\Lambda(1405)$, $N(1535)$, $\Lambda(1670)$, 
$\Sigma(1620)$ and $\Xi(1620)$
as dynamically generated objects. They used the
dimensional regularization scheme with channel-dependent 
subtraction constants, $a_i$.
In particular, 
the subtraction constants in $S=0$
depended significantly on the channel,
while, as reported in Ref.~\citen{Oller:2000fj},
it was found that
a common 
subtraction constant in the $S=-1$ channel reproduces
the total cross sections of the $K^- p$ scattering as well as
the $\Lambda(1405)$ properties.
Note also that in a similar model with a different regularization
scheme~\cite{Garcia-Recio:2003ks}, the position of the poles and their
properties are changed.

\begin{figure}[tbp]
    \centering
    \includegraphics[width=10cm,clip]{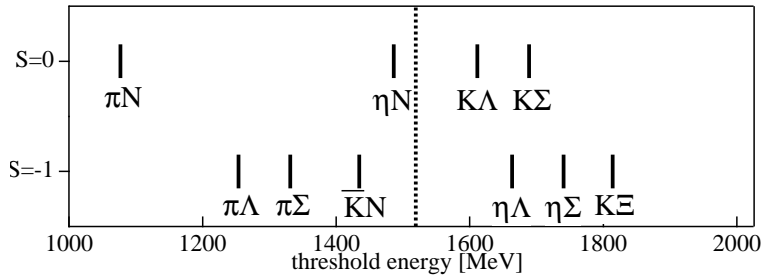}
    \caption{\label{fig:massdif}
    Threshold energies of 
    meson-baryon scattering in the $S=-1$ and $S=0$ channels.
    The dotted line in the middle represents
    the averaged energy of 
    all the meson-baryon thresholds.}
\end{figure}

In this work, we raise the question of whether or not
such a channel dependence of subtraction constants
could be dictated by the flavor $SU(3)$ breaking effects of
an underlying theory.
As we will discuss in detail, it is shown that 
the subtraction constants should not depend on the scattering 
channel in the $SU(3)$ limit~\cite{Jido:2003cb,Hyodo:2002pk}. 
The $SU(3)$ breaking should have a
significant effect on the observed quantities.
This is expected from, for instance, the large dependence
of the threshold energies on the meson-baryon channels,
as shown in Fig.~\ref{fig:massdif}.
This is particularly true for $S=0$,
in which case the lowest threshold energy of the
$\pi N$ channel deviates considerably
from the mean value.
Furthermore, it was discussed in Ref.~\citen{Kolomeitsev:2003kt}
that the number of channel-dependent subtraction constants
for all $SU(3)$ channels exceeds the number of available
counter terms of chiral order $p^3$.

In order to study the above questions, we consider 
the following two cases:
\begin{itemize}
    \item  We use a common subtraction constant for
	   all scattering channels and determine whether this simplified
	   calculation works.

    \item  When this method does not work,
	   we introduce the flavor
	   $SU(3)$ breaking effects into the interaction kernel.
\end{itemize}
In this way, we expect that the parameters in previous 
treatments 
could be controlled
based on appropriate physical considerations.
This would allow us to extend this method to other channels 
with predictive power.
Note that the use of a single subtraction constant was first examined in
Ref.~\citen{Lutz:2000us}.
In this work, we concentrate on
$s$ wave scattering, because the $p$ wave contribution
to the total cross sections was shown to be small in the $S=-1$ 
channel in Ref.~\citen{Jido:2002zk}.

This paper contains a detailed study of the results in
Ref.~\citen{Hyodo:2002pk}.
In \S~\ref{sec:formulation},
we present the formulation of the chiral unitary model.
The calculation with a common subtraction constant
and comparison with the results of previous works are
given in \S~\ref{sec:common}.
We then introduce the flavor $SU(3)$ breaking effects
in the interaction kernel
and present numerical results
in \S~\ref{sec:breaking}.
We discuss the results
and summarize this work
in \S~\ref{sec:discussion}.

\section{Formulation}\label{sec:formulation}

In this section we briefly review the formulation of
the chiral unitary model.
We derive the basic interaction of meson-baryon scattering
from the lowest-order chiral Lagrangian,
and we maintain the unitarity of the S-matrix.
There are several methods
that recover the unitarity of the S-matrix such as 
solving the Bethe-Salpeter equation (BSE)~\cite{Oset:1998it},
the inverse amplitude method (IAM)~\cite{GomezNicola:2000wk},
the N/D method~\cite{Oller:2000fj}, and so on.
In this work, we adopt the N/D method~\cite{Chew:1960iv},
because this method provides a general form of the T-matrix
using the dispersion relation
and the analyticity of the inverse of the T-matrix.
Recently, the N/D method has been applied to coupled
channel meson-baryon scattering~\cite{Meissner:1999vr,Oller:2000fj}.
It was found that 
the final form of the T-matrix derived from the N/D method 
is essentially equivalent to the result given in Ref.~\citen{Oset:1998it}
derived from the BSE.

The chiral Lagrangian for  baryons in the
lowest-order of the chiral expansion is given by~\cite{Donoghue:1992dd}
\begin{equation}
    \mathcal{L}_{\text{lowest}}
    =\Tr\Bigl(\bar{B}(i\Slash{\mathcal{D}}-M_{0})B
    -D(\bar{B}\gamma^{\mu}\gamma_{5}
    \{A_{\mu},B\})
    -F(\bar{B}\gamma^{\mu}\gamma_{5}
    [A_{\mu},B])\Bigr) \ .
    \label{eq:lowestLag}
\end{equation}
Here $D$ and $F$ are coupling constants.
In Eq.~\eqref{eq:lowestLag},
the covariant derivative $\mathcal{D}_\mu$,
the vector current $V_{\mu}$, 
the axial vector current $A_{\mu}$ and
the chiral field $\xi$ are defined by
\begin{align}
    \mathcal{D}_{\mu}B =& \partial_{\mu}B+i[V_{\mu},B]\ , \\
    V_{\mu} =&-\frac{i}{2}(\xi^{\dag}\partial_{\mu}\xi
    +\xi\partial_{\mu}\xi^{\dag})\ ,  \\
    A_{\mu}=&-\frac{i}{2}(\xi^{\dag}\partial_{\mu}\xi
    -\xi\partial_{\mu}\xi^{\dag})\ , \\
    \xi(\Phi)  =&\exp\{i\Phi/\sqrt{2}f\} \ , 
\end{align}
where $f$ is the meson decay constant; here we take an averaged value
$f = 1.15 f_\pi$ with $f_{\pi}=93$ MeV. 
The meson and baryon fields are expressed in $SU(3)$ matrix form as
\begin{eqnarray}
    B&=&\begin{pmatrix}
    \frac{1}{\sqrt{2}}\Sigma^{0}+\frac{1}{\sqrt{6}}\Lambda & 
    \Sigma^{+} & p \\
    \Sigma^{-} & -\frac{1}{\sqrt{2}}\Sigma^{0}
    +\frac{1}{\sqrt{6}}\Lambda & n \\
    \Xi^{-} & \Xi^{0} & -\frac{2}{\sqrt{6}}\Lambda
    \end{pmatrix} \ , \\
    \Phi&=&\begin{pmatrix}
    \frac{1}{\sqrt{2}}\pi^{0}+\frac{1}{\sqrt{6}}\eta & 
    \pi^{+} & K^{+} \\
    \pi^{-} & -\frac{1}{\sqrt{2}}\pi^{0}
    +\frac{1}{\sqrt{6}}\eta & K^{0} \\
    K^{-} & \bar{K}^{0} & -\frac{2}{\sqrt{6}}\eta .
    \end{pmatrix} \ .
\end{eqnarray}
In the Lagrangian \eqref{eq:lowestLag}, $M_0$ denotes
the common mass of the octet baryons. 
However, we use the observed values of the baryon masses
in the following calculations.
The mass splitting among the octet baryons in the Lagrangian level
are introduced consistently with the $SU(3)$ breaking
in \S~\ref{sec:breaking}.

\begin{figure}[tbp]
    \centering
    \includegraphics[width=5cm,clip]{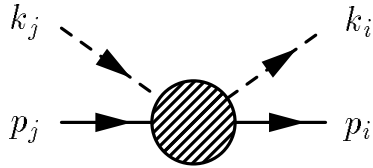}
    \caption{Definition of the momentum variables.
    The dashed and solid lines represent mesons
    and baryons, respectively.}
    \label{fig:kinematics}
\end{figure}

The $s$ wave interactions at tree level come from  
the Weinberg-Tomozawa (WT) interaction,
which is in the vector coupling term in the covariant derivative:
\begin{equation}
    \mathcal{L}_{WT}
    =\Tr\Bigl(\bar{B}i\gamma^{\mu}\frac{1}{4f^{2}}
    \bigl[(\Phi\partial_{\mu}\Phi
    -\partial_{\mu}\Phi\Phi),
    B\bigr]\Bigr) \ .
    \label{eq:WTLag}
\end{equation}
From this Lagrangian, the meson-baryon 
scattering amplitude at tree level is given by
\begin{align}
    V_{ij}^{(WT)}=&-\frac{C_{ij}}{4f^{2}}
    \bar{u}(p_i)(\Slash{k}_i+\Slash{k}_j)u(p_j)
    \notag \\
    =&-\frac{C_{ij}}{4f^{2}}
    (2\sqrt{s}-M_i-M_j)\sqrt{\frac{E_i+M_i}{2M_i}}
    \sqrt{\frac{E_j+M_j}{2M_j}},
    \label{eq:WTint} 
\end{align}
where the indices $(i,j)$ denote the channels
of the meson-baryon scattering, and 
$M_i$ and $E_i$ are the mass and the energy
of the baryon in the channel $i$, respectively.
These masses and factors come from the spinors
of the baryons.
It seems reasonable
to use the common mass $M_0$ in the Lagrangian
as in Ref.~\citen{Oller:2000fj}.
However,
in this paper 
we adopt the physical masses,
as in Refs.~\citen{Oset:2001cn,Inoue:2001ip,Ramos:2002xh}.
Indeed, we have checked that
the results obtained with a common mass are qualitatively 
similar to the results obtained with observed masses.
The channels $(i,j)$ are shown in Table~\ref{tab:classification}
of the Appendix.
The kinematics of this vertex are depicted in
Fig.~\ref{fig:kinematics},
and $s$ in Eq.~\eqref{eq:WTint} is defined as $s=(k+p)^{2}$.
The last line is obtained in the center of mass frame
with nonrelativistic reduction.
The coefficient $C_{ij}$ is fixed by chiral symmetry,
and the explicit form of $C_{ij}$ is given in Ref.~\citen{Oset:1998it}
for $S=-1$ and in Ref.~\citen{Inoue:2001ip} for $S=0$.

In the coupled channel formulation, the T-matrix takes a matrix form.
The unitarity condition is guaranteed by the optical theorem, 
\textit{i.e.}
$-2\im[T_{ii}]=T_{ik}\rho_{k} T^{*}_{ki}$,
which can be written as 
\begin{align}
    2\im [T^{-1}_{ii}] = \rho_{i} \ ,
    \label{eq:optical}
\end{align}
where the normalization of the $T$-matrix is defined by
\begin{equation}
    S_{ij}=1-i\left(\frac{\sqrt{2M_i |\bm{q}_{i}| 2M_j 
    |\bm{q}_{j}|}}{
    4\pi\sqrt{s}}\right)T_{ij} \ .
    \nonumber
\end{equation}
With the condition \eqref{eq:optical} and
the dispersion relation for $T^{-1}_{ii}$,
we find a general form of the T-matrix
using the N/D method.
Following Ref.~\citen{Oller:2000fj}, we write
\begin{equation}
    T^{-1}_{ij}(\sqrt{s})
    =\delta_{ij}\Bigl(\tilde{a}_i(s_0)+\frac{s-s_0}{2\pi}
    \int_{s^{+}_i}^{\infty}ds^{\prime}
    \frac{\rho_i(s^{\prime})}{(s^{\prime}-s)(s^{\prime}-s_0)}
    \Bigr)
    +\mathcal{T}^{-1}_{ij} \ ,
    \label{eq:NDamplitude}
\end{equation}
where $s^{+}_i$ is the value of $s$ at the threshold of the channel $i$,
and $s_0$ is the subtraction point.
The parameter $\tilde{a}_i(s_0)$
is the subtraction constant
and is a free parameter within the N/D method. The matrix
$\mathcal{T}_{ij}$ is determined by the chiral perturbation theory,
as discussed below. 
In the derivation of Eq.~\eqref{eq:NDamplitude}, 
we have ignored the left-hand cuts,
which correspond to $u$-channel diagrams of the crossing symmetry.

\begin{figure}[tbp]
    \centering
    \includegraphics[width=10cm,clip]{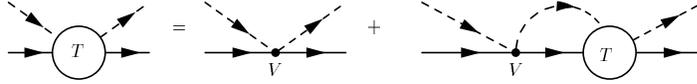}
    \caption{Diagrammatic interpretation of Eq.~\eqref{eq:result}.}
    \label{fig:BSE}
\end{figure}%

Let us assume that the intermediate states of the meson-baryon
scattering are composed of one octet meson and one octet baryon.
We do not consider the case of multiple mesons and excited baryons, 
such as
$\pi\pi N$ and $\pi \Delta$.  
In this case, the phase space $\rho_i$ in Eq.~\eqref{eq:optical} is
written
\begin{equation}
    \rho_{i}(\sqrt{s})=\frac{2M_i|\bm{q}_{i}|}{4\pi\sqrt{s}},
    \label{eq:rho}
\end{equation}
where $\bm{q}_{i}$ is a three-momentum of the intermediate meson
on the mass shell.
Let us define the $G$ function by
\begin{equation}
    G_{i}(\sqrt{s})=-\tilde{a}_i(s_0)
    -\frac{s-s_0}{2\pi}
    \int_{s^{+}_i}^{\infty}ds^{\prime}
    \frac{\rho_i(s^{\prime})}{(s^{\prime}-s)(s^{\prime}-s_0)} \ ,
    \label{eq:gfunction}
\end{equation}
which takes the same form as, up to a constant,
the ordinary meson-baryon loop function:
\begin{align}
    G_i(\sqrt{s})
    &= i\int\frac{d^{4}q}{(2\pi)^{4}}
    \frac{2M_i}{(P-q)^{2}-M_i^{2}+i\epsilon}
    \frac{1}{q^{2}-m_i^{2}+i\epsilon} \ .
\end{align}
This integral should be regularized with
an appropriate regularization scheme. 
In the dimensional regularization,
the integral is calculated as 
\begin{equation}
    \begin{split}
    G_i(\sqrt{s})&=\frac{2M_{i}}{(4\pi)^{2}}
    \Biggl\{a_i(\mu)+\ln\frac{M_i^{2}}{\mu^{2}}
    +\frac{m_i^{2}-M_i^{2}+s}{2s}\ln\frac{m_i^{2}}{M^{2}_i} \\
    &\quad\quad+\frac{\bar{q}_i}{\sqrt{s}}
    \Bigl[\ln(s-(M^{2}_i-m_i^{2})+2\sqrt{s}\bar{q}_i)
    +\ln(s+(M^{2}_i-m_i^{2})+2\sqrt{s}\bar{q}_i) \\
    &\quad\quad
    -\ln(-s+(M^{2}_i-m_i^{2})+2\sqrt{s}\bar{q}_i)
    -\ln(-s-(M^{2}_i-m_i^{2})+2\sqrt{s}\bar{q}_i)
    \Bigr]\Biggl\},
    \end{split}
    \label{eq:loop}
\end{equation}
where $\mu$ is the regularization scale,
$a_i$ is the subtraction constant,
and $\bar{q}_i$ is defined by
\begin{equation}
    \bar{q}_{i}(\sqrt{s})=\frac{\sqrt{(s-(M_i-m_i)^{2})
    (s-(M_i+m_i)^{2})}}{2\sqrt{s}} \ .
    \label{eq:qdef}
\end{equation}
In the tree approximation,
only the $\mathcal{T}_{ij}$ term survives
in Eq.~\eqref{eq:NDamplitude}.
This may be identified with the WT interaction
$V^{(WT)}$ in Eq.~\eqref{eq:WTint}. 
Therefore, the resulting T-matrix is written
\begin{align}
    T^{-1}&=-G+(V^{(WT)})^{-1} \ , \\ 
    T&=V^{(WT)}+
    V^{(WT)}GT \ .
    \label{eq:result}    
\end{align}
This is the algebraic equation for the T-matrix,
which corresponds to the integral BSE.
The diagrammatic interpretation of Eq.~\eqref{eq:result}
is displayed in Fig.~\ref{fig:BSE}.

The subtraction constants $a_i(\mu)$ in Eq.~\eqref{eq:loop},
in principle, 
would be related to the counter terms in the higher-order Lagrangian
in the chiral perturbation theory.
In previous works~\cite{Oset:2001cn,Inoue:2001ip},
the subtraction constants $a_i$ were fitted
using the data for
$\bar{K}N(S=-1)$ and $\pi N(S=0)$ scatterings.
In Table~\ref{tbl:subtractions}, 
we list the subtraction constants 
used in Refs.~\citen{Oset:2001cn} and \citen{Inoue:2001ip}.
In the table, in order to compare the channel dependence
of the subtraction constants,
we take the regularization scale at 
$\mu = 630$ MeV in the both channels.
Changing the regularization scale,
the subtraction constants are simply shifted by 
$a(\mu^{\prime})=a(\mu)+2\ln(\mu^{\prime}/\mu)$.
From this table, we see that the values of $a_i$ values for $S=0$
differ significantly.
In the rest of this paper
we refer to these parameters as the ``channel-dependent $a_i$''.

\begin{table}[tbp]
    \centering
    \begin{tabular}{|ccccccc|} \hline
    \multicolumn{7}{|c|}{channel-dependent $a_i$ ($S=-1$)} \\
    \hline
    channel & $\bar{K}N$ & $\pi\Sigma$ & $\pi\Lambda$
    & $\eta\Lambda$ & $\eta\Sigma$ & $K\Xi$  \\
    \hline
    $a_i$& $-1.84$ & $-2.00$ & $-1.83$
    & $-2.25$ & $-2.38$ & $-2.67$  \\
    \hline
    \end{tabular}
   \vspace{0.5cm}
   \begin{tabular}{|ccccc|} \hline
   \multicolumn{5}{|c|}{channel-dependent $a_i$ ($S=0$)}  \\
    \hline
    channel & $\pi N$ & $\eta N$ & $K\Lambda$ & $K\Sigma$  \\
    \hline
    $a_i$ & 0.711 & $-1.09$ & 0.311 & $-4.09$  \\
    \hline
    \end{tabular}
    \caption{Channel-dependent subtraction constants  $a_i$ used in
    Refs.~\citen{Oset:2001cn} and \citen{Inoue:2001ip}
    with the regularization scale $\mu=630$ MeV. 
   For the $S=0$ channel, although the original values of $a_i$
   are obtained with $\mu=1200$ MeV,
   here we give the values of $a_i$ corresponding 
    to $\mu=630$ MeV obtained using the relation 
    $a(\mu^{\prime})=a(\mu)+2\ln(\mu^{\prime}/\mu)$.}
    \label{tbl:subtractions}
\end{table}%

\section{Calculation with a common subtraction constant
}\label{sec:common}

In this section, we present calculations in which a single subtraction 
constant $a$ is commonly used
in the meson-baryon loop function \eqref{eq:loop}
in order to determine the role of the channel-dependent $a_i$
in reproducing the observed cross sections and the resonance
properties. 
A channel-independent regularization scheme was first used in
Ref.~\citen{Lutz:2000us}.

Let us first show thatÊ in the $SU(3)$ limit, together with the 
constraint in the chiral unitary model, there is only
one subtraction constant~\cite{Jido:2003cb,Hyodo:2002pk}. 
Under $SU(3)$ symmetry, the scattering amplitudes of one octet meson and 
one octet baryon are composed of $SU(3)$ irreducible 
representations.ÊÊ The amplitudes satisfy the following scattering 
equation in each representation:
\begin{equation} 
T(D) = V(D) + V(D) G(D)T(D) \ .
\end{equation} 
Here, $D$ represents an $SU(3)$ irreducible representation, 
$D = 1, 8, 8, 10, \bar{10}$ and 27. 
Therefore, on one hand, 
the functions $G$, or equivalently the subtraction constants $a_i$,
are represented by diagonal matrices in the $SU(3)$ basis. 
On the other hand, because $G$ functions are given as 
loop integrals, as shown in (14) and (15), they are also diagonal 
in the particle basis ($\pi^{-} p, \eta \Lambda, \cdots$). 
These observations imply that the subtraction constants
are components of a diagonal matrix both in $SU(3)$ and in particle bases, 
whichÊ are transformed uniquely with
a unitary matrix of $SU(3)$ Clebsch-Gordan 
coefficients,
\begin{equation} 
a(D) = \sum_kÊ U_{Dk} a_k (U^\dagger)_{kD} \ .
\end{equation} 
This can happen when the subtraction constants 
are proportional to unity. 
Hence, the subtraction constants are not dependent 
on the channel in the $SU(3)$ limit.

Now, we discuss the case $S=-1$,
in which the subtraction constants $a_i$ do not depend strongly 
on the channel, as shown in Table \ref{tbl:subtractions}. 
Therefore, it is expected that a calculation with
a common value $a$ gives a good description
if we choose a suitable value.

Next we study the $S=0$ channel using a common 
subtraction constant.
Here, we find that common value $a$ 
cannot simultaneously reproduce 
the resonance properties and the $S_{11}$ amplitude
in the low energy region.

In order to concentrate on the role of the subtraction constants 
and to deduce the channel dependence, 
we make the following simplifications for the calculations of 
the $S=-1$ and $S=0$ channels:
\begin{itemize}
    \item  We use an averaged value for the meson decay constants,
    $f=1.15f_\pi=106.95$ MeV, while in Ref.~\citen{Inoue:2001ip},
    physical values were taken as $f_{\pi}=93\text{ MeV},\;
    f_{K}=1.22f_{\pi},\;f_{\eta}=1.3f_{\pi}$.

    \item  We do not include the effect of vector meson exchanges
    and $\pi\pi N$ channels to reproduce the $\Delta(1620)$ resonance,
    which were considered in Ref.~\citen{Inoue:2001ip}.
\end{itemize}
With these simplifications, the calculations in
the $S=-1$ and $S=0$ channels are based on exactly the same formulation;
the differences are in the flavor $SU(3)$ coefficients $C_{ij}$
in Eq.~\eqref{eq:WTint} and in the channel-dependent subtraction constants.

\subsection{The $S=-1$ channel ($\bar{K} N$ scattering)}

In the $S=-1$ channel, the subtraction constants $a_i$
obtained in Ref.~\citen{Oset:2001cn} do not
depend strongly on the channel,
as shown in Table \ref{tbl:subtractions}. 
In Ref.~\citen{Oller:2000fj}, a common value of $a\sim -2$
was used.
This value was ``naturally" obtained from matching with 
the three-momentum cut-off regularization with $\Lambda=630$ MeV. 
In both works,
the total cross sections of the $K^- p$ 
scattering and the 
mass distribution of the $\pi \Sigma$ channel with $I=0$, where the 
$\Lambda(1405)$ resonance is seen,
were reproduced very well.
In Ref.~\citen{Oset:2001cn}, the 
$\Lambda(1670)$ resonance was also obtained
with the channel-dependent subtraction constants,
and its properties were investigated by analyzing the speed plots
in the $I=0$ channels. 

Here we search for one common value $a$
to be used in all channels for $S=-1$.
In order to fix this common value $a$, 
we use threshold properties of the $\bar K N$ scattering,
which are well observed in the branching 
ratios~\cite{Nowak:1978au,Tovee:1971ga}:
\begin{align}
    \gamma=&\frac{\Gamma(K^{-}p\to\pi^{+}\Sigma^{-})}
    {\Gamma(K^{-}p\to\pi^{-}\Sigma^{+})}\sim  2.36\pm0.04 \
    ,\nonumber \\
    R_c=&\frac{\Gamma(K^{-}p\to\text{charged particles})}
    {\Gamma(K^{-}p\to\text{all})}\sim 0.664\pm 0.011\ , \nonumber \\
    R_n=&\frac{\Gamma(K^{-}p\to\pi^{0}\Lambda)}
    {\Gamma(K^{-}p\to\text{neutral particles})}\sim 0.189\pm 0.015 \ .
    \label{eq:branch}
\end{align}
After fitting, we find the optimal value $a=-1.96$, 
with which the threshold branching ratios are obtained, as shown
in Table \ref{tbl:branch}. 
The result obtained using the common value $a=-1.96$
does not differ much from that obtained with channel-dependent values,
and also the value $a=-1.96$ is close to the averaged value
of the channel-dependent subtraction constants $a_i$, namely $\sim -2.15$. 
Therefore, the threshold properties are not sensitive to such a 
fine tuning of the subtraction constants.

\begin{figure}[tbp]
    \centering
    \includegraphics[width=8.3cm,clip]{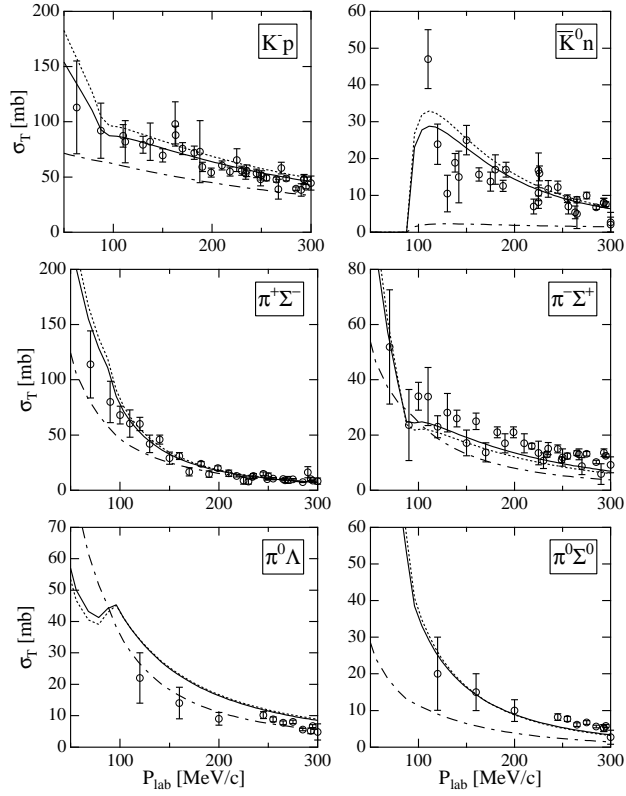}
    \caption{Total cross sections of 
    $K^{-}p$ scattering ($S=-1$)
    as functions of  $P_{\text{lab}}$,
    the three-momentum of the initial $K^{-}$
    in the laboratory frame.
    The dotted curve represent 
    the results obtained with
    the channel-dependent $a_i$,
    the solid curves represent the results obtained 
    with the common value $a=-1.96$,
    and the dash-dotted curves represent the results
    obtained
    with the common value $a=-2.6$.
    The open circles with error bars are experimental data
    taken from Refs.~\citen{Mast:1976pv,
    Ciborowski:1982et,Bangerter:1981px,
    Mast:1975sx,Sakitt:1965kh,
    PR131.2248,PRL14.29,PRL8.23,PL16.89,PR123.2168,PL21.349,NC16.848}.}
    \label{fig:S-1commoncross}\vspace{0.5cm}
\end{figure}

\begin{figure}[tbp]
    \centering
    \includegraphics[width=8.3cm,clip]{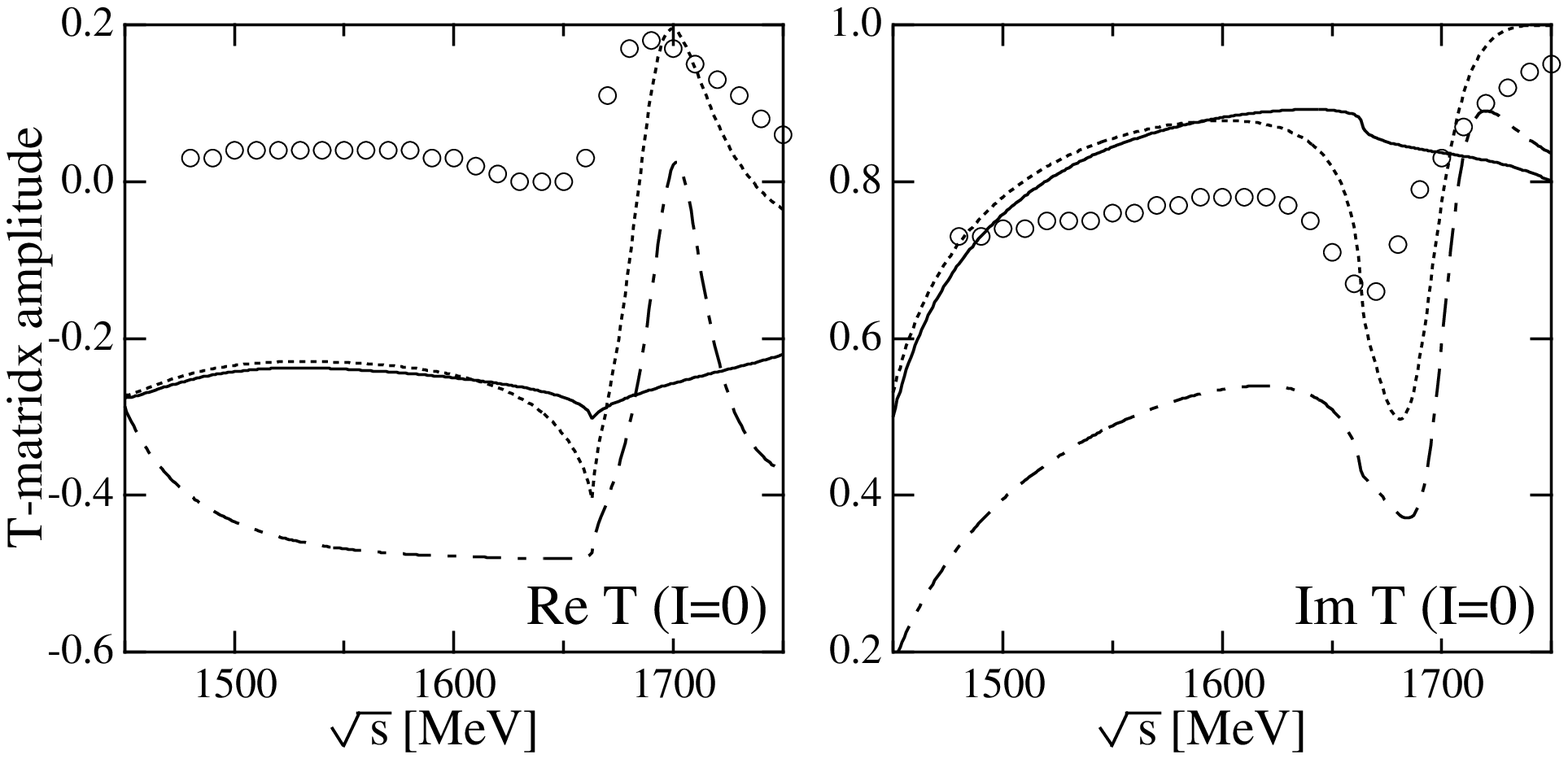}
    \caption{Real and imaginary parts of the 
    T-matrix amplitude of $\bar KN\to\bar KN$ with $I=0$.
    The dotted curves represent the results obtained with
    the channel-dependent $a_i$,
    the solid curves represent the results obtained
    with the common value $a=-1.96$,
    and the dash-dotted curves represent the results obtained
    with the common value $a=-2.6$.
    The open circles 
    are experimental data
    taken from Ref.~\citen{Gopal:1977gs}.}
    \label{fig:S-1commontmat}\vspace{0.5cm}
    \centering
    \includegraphics[width=8.3cm,clip]{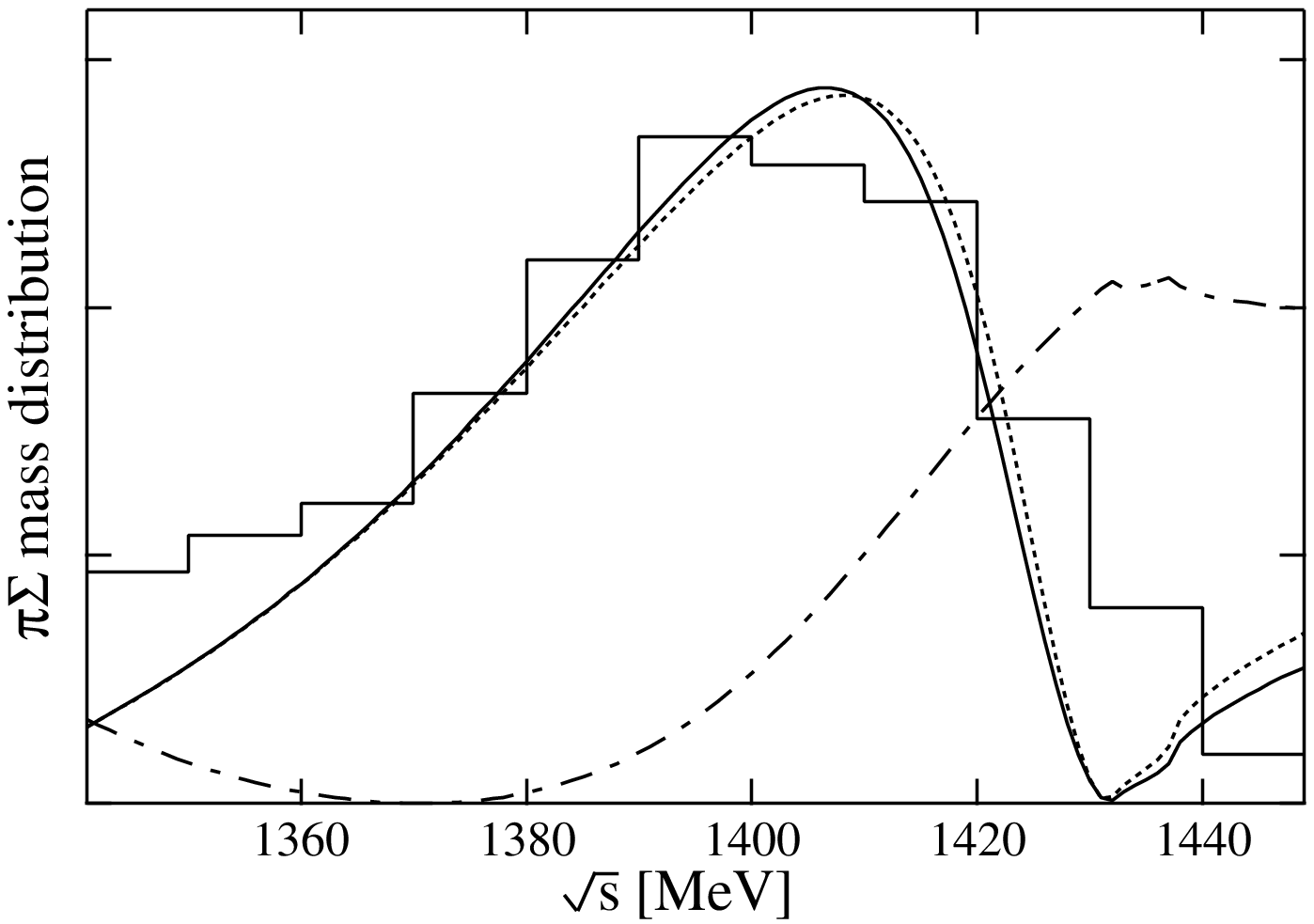}
    \caption{The mass distributions of the $\pi \Sigma$ channel
    with $I=0$.
    The dotted curve represents the result obtained with
    the channel-dependent $a_i$,
    the solid curve represents the result obtained 
    with the common value $a=-1.96$,
    and the dash-dotted curve represents
    the result obtained with the common value $a=-2.6$.
    The histogram represents experimental data
    taken from Ref.~\citen{Hemingway:1985pz}.}
    \label{fig:S-1commonmdist}
\end{figure}

Using the common value $a=-1.96$, we calculate 
the total cross sections of the $K^{-}p $ scattering
(Fig.~\ref{fig:S-1commoncross}, solid curves),
the T-matrix amplitude of the $\bar KN$ scattering
with $I=0$
(Fig.~\ref{fig:S-1commontmat}, solid curves),
and the mass distributions of the $\pi \Sigma$ channel with $I=0$ 
(Fig.~\ref{fig:S-1commonmdist}, solid curves).
We also plot the results obtained 
with the channel-dependent $a_i$ from the calculation given
in Ref.~\citen{Oset:2001cn} in
Figs.~\ref{fig:S-1commoncross}, \ref{fig:S-1commontmat}
and \ref{fig:S-1commonmdist} as the
dotted curves. Here, we find that the present calculations 
give results that are slightly different from those of the calculations
with the channel-dependent $a_i$ 
in the total cross sections and the $\pi \Sigma$ mass distributions.
Therefore, the $\Lambda (1405)$ resonance is well reproduced
with the common value $a=-1.96$,
which is consistent with the results in Ref.~\citen{Oller:2000fj}. 
However, the resonance $\Lambda(1670)$ disappears when this
common value $a$ is used, as we see in the T-matrix amplitude of $\bar KN
\rightarrow \bar KN$ with $I=0$ in Fig.~\ref{fig:S-1commontmat}.
As pointed out in Ref.~\citen{Oset:2001cn},
the $\Lambda(1670)$ resonance structure is
very sensitive to the value of $a_{K\Xi}$.
Indeed, we have checked that the $\Lambda(1670)$ resonance is reproduced
when we choose $a_{K\Xi}\sim -2.6$
with the other $a_i$ unchanged, \textit{i.e.}, at $-1.96$.
In a recent publication, it was shown that the poles of
$\Lambda(1405)$ and $\Lambda(1670)$ are simultaneously
 reproduced by taking into
account the approximate crossing symmetry without considering
explicitly the channel dependence~\cite{Garcia-Recio:2003ks}.
The inclusion of the crossing symmetry is, however, beyond the
scope of the present discussion.

If we choose  $a=-2.6$ for all subtraction constants,
the threshold branching ratios are obtained as 
$\gamma= 2.41$, $R_c  = 0.596$ and $R_n = 0.759$, and
the agreement with the experimental data becomes poor, as shown
in Figs.~\ref{fig:S-1commoncross} and \ref{fig:S-1commontmat}.
In particular, the $K^{-}p\to\bar{K}^{0}n$ cross section is underestimated,
and also the resonance structure of $\Lambda(1405)$
disappears in the $\pi\Sigma$ mass distribution 
(Fig.~\ref{fig:S-1commonmdist}).
As we change all subtraction constants from $a=-1.96$ to $a=-2.6$ gradually,
the position of the peak of $\Lambda(1405)$ moves to the lower energy side
and finally disappears under the $\pi \Sigma$ threshold.  Therefore, 
using the common value $a\sim -2$ is essential
to reproduce the resonance properties of 
$\Lambda(1405)$ and the total cross sections of the $K^- p$ scattering
in the low energy region.

\begin{table}[tbp]
     \centering
     \begin{tabular}{|c|ccc|}\hline
	  & $\gamma$ & $R_c$ & $R_n$  \\
	 \hline
	 experiment  & $2.36\pm0.04$ & $0.664\pm 0.011$
	& $0.189\pm 0.015$  \\
	 \hline
	 channel-dependent $a_i$ & 1.73  & 0.629 & 0.195  \\
	 common value $a$ & 1.80  & 0.624 & 0.225  \\
	 $SU(3)$ breaking & 2.19  & 0.623 & 0.179  \\
	 \hline
     \end{tabular}
     \caption{Threshold branching ratios calculated 
     with channel-dependent $a_i$,
     common value $a=-1.96$, and
     $a=-1.59$ with the $SU(3)$ breaking interaction.
     The experimental values were taken from
     Refs.~\citen{Nowak:1978au} and \citen{Tovee:1971ga}.}
     \label{tbl:branch}
\end{table}%



\subsection{The $S=0$ channel ($\pi N$ scattering)}

\begin{figure}[tbp]
    \centering
    \includegraphics[width=8.3cm,keepaspectratio,clip]{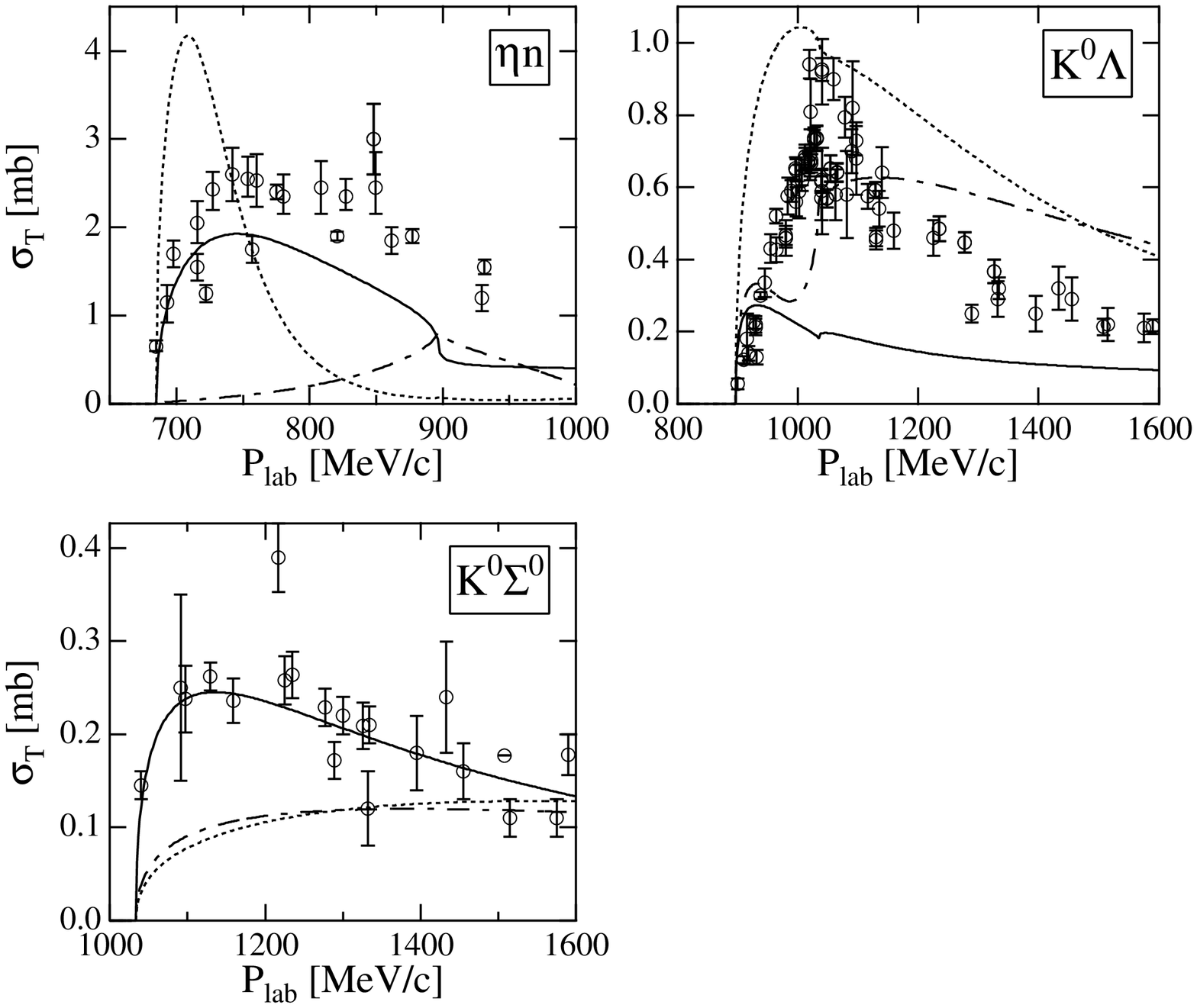}
    \caption{Total cross sections of 
    $\pi^{-} p$ scattering ($S=0$)
    as functions of  $P_{\text{lab}}$,
    the three-momentum of the initial $\pi^{-}$
    in the laboratory frame.
    The dotted curves represent
     the results obtained with channel-dependent $a_i$,
    the dash-dotted curves represent
     the results obtained with the common value $a=-1.96$,
     obtained for $S=-1$,
    and the solid curves represent
     the results obtained with the common value $a=0.53$.
    The open circles with error bars are the experimental data
    taken from Refs.~\citen{Batinic:1995kr,Hart:1980jx,Saxon:1980xu,
    Baker:1978bb,Baker:1978qm,
    Nelson:1973py,Thomas:1973uh,
    Jones:1971zm,Binford:1969ts,VanDyck:1969ay,
    PRD11.1,PR155.1455,PR133.B457,PR132.1778,
    PR109.1358,PRL8.332,PRL3.394,
    NC53A.745,NC42A.606,NC10.468}.}
    \label{fig:S0commoncross} \vspace{0.5cm}
    \centering
    \includegraphics[width=8.3cm,clip]{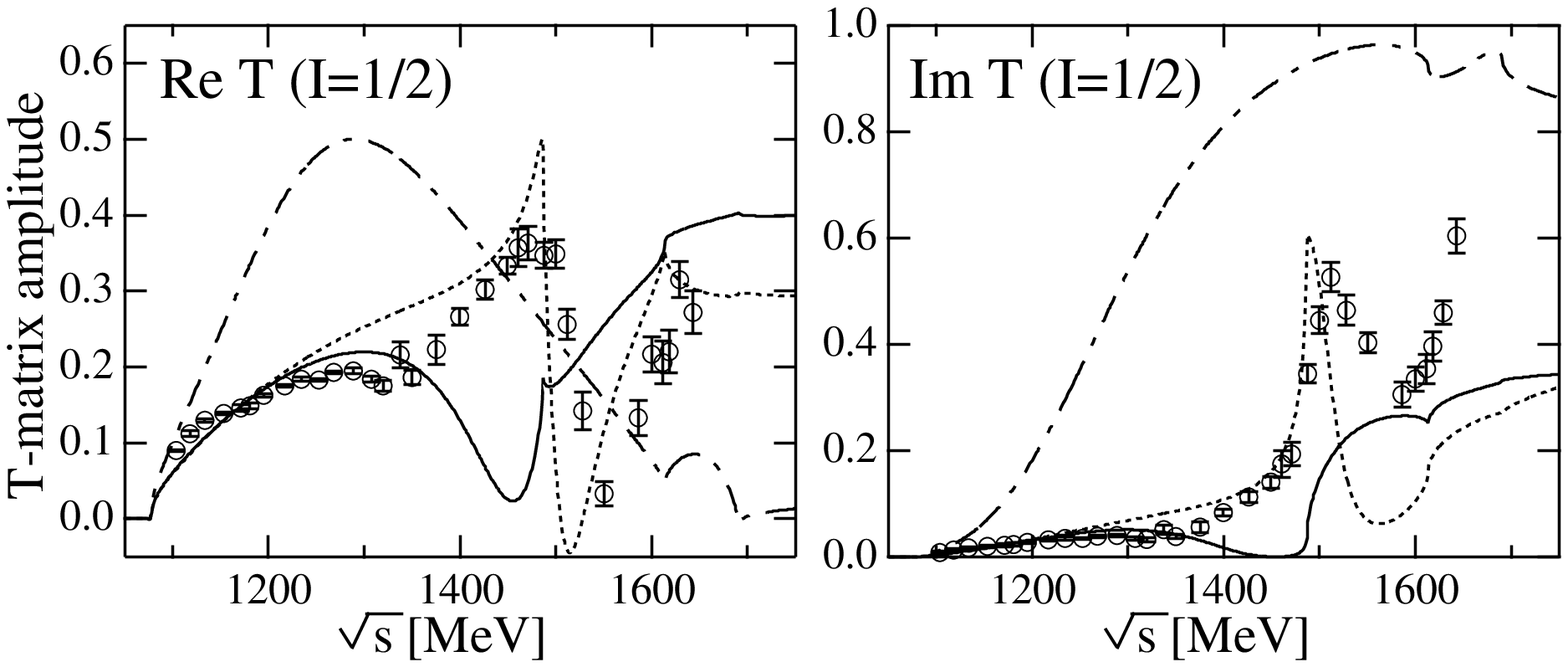}
    \caption{Real and imaginary parts of the $S_{11}$
    T-matrix amplitudes of $\pi N\to \pi N$.
    The dotted curves represent
     the results obtained with channel-dependent $a_i$,
    the dash-dotted curves represent
     the results obtained with the common value $a=-1.96$,
     obtained for $S=-1$,
    and the solid curves show
     the results with the common value $a=0.53$.
    The open circles with error bars are experimental data
    taken from Ref.~\citen{CNS}.}
    \label{fig:S0commontmat}
\end{figure}

In Ref.~\citen{Inoue:2001ip}, the total cross sections of the $\pi^- p$ 
inelastic scattering and the resonance properties of 
the $N(1535)$ were reproduced
well by using channel-dependent $a_i$.
After the simplification applied to $f$ and inelastic channels,
the agreement with the data is still acceptable,
as shown in Figs.~\ref{fig:S0commoncross} and 
\ref{fig:S0commontmat} by the dotted curves,
as long as channel-dependent $a_i$ are employed.
In the T-matrix elements of the  $\pi N$ scattering in the $S_{11}$ channel, 
we see a kink structure around the energy $\sqrt{s}\sim 1500$ MeV,
which corresponds to the $N(1535)$ resonance~\cite{Inoue:2001ip}.

In the previous subsection, we obtained the common subtraction constant
$a=-1.96$ with which the $\bar KN$ total cross sections 
and the $\Lambda(1405)$ properties are reproduced well.
First, we use this common value of $a$ for the $S=0$ channel.
It is worth noting that in Ref.~\citen{Garcia-Recio:2003ks}, 
$N(1535)$ and $\Lambda(1405)$ were reproduced
with the channel-independent renormalization scheme.
Shown in Figs.~\ref{fig:S0commoncross} 
and \ref{fig:S0commontmat} by the dash-dotted curves
are the results with $a=-1.96$ for
the total cross sections of the $\pi^- p \rightarrow \pi^0 \eta$, 
$K^0 \Lambda$ and $K^0 \Sigma$ scatterings, and the $S_{11}$ 
T-matrix amplitude of $\pi N \rightarrow \pi N$. 
As can be seen in Figs.~\ref{fig:S0commoncross} and 
\ref{fig:S0commontmat}, the results with $a=-1.96$
in the $S=0$ channel are far from the experimental data.
In particular, in the $\pi^-  p \rightarrow \eta n$ cross section,
the threshold behavior disagrees with the experiment, 
and a resonance structure of $N(1535)$ disappears. 
In addition, as shown in Fig.~\ref{fig:S0commontmat},
the T-matrix amplitude of the $S_{11}$ channel is overestimated, 
and an unexpected resonance has been generated 
near $\sqrt s \sim 1250$ MeV.

Next, we search a single optimal subtraction constant
within the $S=0$ channel,
because an unnecessary resonance is obtained with $a=-1.96$ at low energy.
In order to avoid the appearance of such an unphysical resonance,
we determine the common subtraction constant $a$ so as to reproduce 
the observed data up to $\sqrt s = 1400$ MeV.
The optimal value is found to be $a=0.53$.
The calculated $S_{11}$ amplitude 
as well as the total cross sections are plotted 
in  Figs.~\ref{fig:S0commoncross} and 
\ref{fig:S0commontmat} by the solid curves.
With this subtraction constant, the low energy behavior of 
the $S_{11}$ amplitude of the $\pi N$ scattering ($\sqrt{s}<1400$ MeV) 
is well reproduced.
Therefore, the scattering length is also reproduced.
However, the $N(1535)$ resonance structure is
not still generated.
We have also checked 
that there is no pole in the scattering amplitudes
in the second Riemann sheet.
Therefore, we conclude that in the $S=0$ channel we cannot reproduce 
simultaneously the $N(1535)$ resonance and the $S_{11}$ amplitude 
at low energy if a single subtraction constant is used
within the present approach.

\section{Flavor $SU(3)$ breaking interactions}\label{sec:breaking}

In previous studies, it has been found that
the channel-dependent subtraction constants
$a_i$ are crucial in order to reproduce
important features of experimental data.
In this section, we consider $SU(3)$ breaking terms of the chiral Lagrangian
in order to see if the channel dependence in the subtraction
constants can be absorbed into those terms.
In this way, we are hoping that the number of free parameters can
be reduced and that the origin of the channel dependence
can be clarified.

\subsection{Flavor $SU(3)$ breaking terms in the chiral Lagrangian}
Here we introduce the flavor $SU(3)$ breaking effects in the chiral 
Lagrangian by the quark masses. 
They are obtained by assuming 
that the current quark mass matrix $\bm{m}$ is 
transformed under the chiral transformation
as $\bm{m} \rightarrow R \bm{m} L^\dagger$
and $\bm{m}^\dagger = \bm{m}$. 
Here we maintain isospin symmetry, that is,
$\bm{m}=\text{diag}(\hat{m},\hat{m},m_s)$.
Then, the $SU(3)$ breaking terms are given
uniquely up to order $\mathcal{O}(m_q)$ as~\cite{Donoghue:1992dd}
\begin{align}
     \begin{split}
	\mathcal{L}_{SB}
	=&-\frac{Z_0}{2}\Tr \Bigl(
	d_m\bar{B}\{\xi\bm{m}\xi
	+\xi^{\dag}\bm{m}\xi^{\dag},B\}
	+f_m\bar{B}[\xi\bm{m}\xi
	+\xi^{\dag}\bm{m}\xi^{\dag},B]
	\Bigr) \\
	&-\frac{Z_1}{2}\Tr(\bar{B}B)
	\Tr(\bm{m}U+U^{\dag}\bm{m}) \ ,
     \end{split}
     \label{eq:SBLag}
\end{align}
where $f_{m}+d_{m}=1$
and $U(\Phi) = \xi^2 =\exp\{i\sqrt{2}\Phi/f\} $.
In this Lagrangian, there are three free parameters,
$Z_0,Z_1, f_m/d_m$,
which are determined by the baryon masses 
and the pion-nucleon sigma term, as we see below.
For the quark mass, we take $m_s/\hat{m}=26$,
which is determined in ChPT from the meson masses.
According to the chiral counting rule, these quark mass terms can be 
regarded as quantities of $\mathcal{O}(p^{2})$, 
if we assume the Gell-Mann--Oakes--Renner
relation~\cite{Gell-Mann:1968rz},
which implies $m_q \propto m_{\pi}^2$.
In this work, we take into account only the terms in Eq.~\eqref{eq:SBLag},
and we do not consider other terms of order $\mathcal{O}(p^{2})$.
We explain the reason in the next subsection.

Expanding the Lagrangian (\ref{eq:SBLag}) in powers of the meson fields,
the zeroth order terms contribute to the baryon mass splitting,
which automatically satisfy the Gell-Mann--Okubo (GMO) mass
formula~\cite{Gell-Mann:1962xb,Okubo:1962jc}.
By using the mass differences among the octet baryons,
we determine the parameters $Z_0$ and $f_m/d_m$.
The $\pi N$ sigma term,
which we take here to be $\sigma_{\pi N} =36.4$ MeV,
is used to determine the parameter $Z_1$.
The resulting parameters are given as
\begin{equation}
Z_0 = 0.528 \ ,  \quad   Z_1=1.56 \ ,    \quad f_m / d_m =-0.31 
\end{equation}
and $M_0=759$ MeV in the Lagrangian~\eqref{eq:lowestLag}.

The meson-baryon interaction Lagrangian with $SU(3)$ breaking is
obtained by picking up the terms with two meson fields.
We find
\begin{align}
     \mathcal{L}_{SB}^{(2)}
     =&\frac{Z_0}{4f^{2}}\Tr\Bigl(
     d_m\bar{B}\bigl\{(2\Phi\bm{m}\Phi
     +\Phi^{2}\bm{m}
     +\bm{m}\Phi^{2}),B\bigr\}
     +f_m\bar{B}\bigl[(2\Phi\bm{m}\Phi
     +\Phi^{2}\bm{m}
     +\bm{m}\Phi^{2}),B\bigr]
     \Bigr)\notag \\
     &+\frac{Z_1}{f^{2}}
     \Tr(\bar{B}B)\Tr(\bm{m}\Phi^{2}) \ .
     \label{eq:MMBBLag}
\end{align}
From this Lagrangian, the basic interaction is given by
\begin{align}
     V^{(SB)}_{ij}=&-\frac{1}{f^{2}}
     \Bigl[
     Z_0\Bigl( (A^{d}_{ij}d_m
     +A^{f}_{ij}f_m )\hat{m}
     +(B^{d}_{ij}d_m
     +B^{f}_{ij}f_m)m_s
     \Bigr) \notag \\*
     &+Z_1\delta_{ij}D^{Z_1}_{i}\Bigr]
     \sqrt{\frac{E_i+M_i}{2M_i}}
     \sqrt{\frac{E_j+M_j}{2M_j}}.
     \label{eq:SBint}
\end{align}
The explicit forms of the coefficients $A_{ij}$,
$B_{ij}$ and $D_i$ are given in the Appendix.
These interaction terms are independent of the meson momenta, unlike
the WT interaction~\eqref{eq:WTint}.

Adding Eq.~\eqref{eq:SBint} to Eq.~\eqref{eq:WTint}
and substituting them into Eq.~\eqref{eq:result},
we obtain the unitarized T-matrix with 
the flavor $SU(3)$ breaking effects as
\begin{equation}
     T=\left[1-\left(V^{(WT)}+V^{(SB)}\right)G\right]^{-1}
     \left(V^{(WT)}+V^{(SB)}\right).
     \label{eq:breakresult}
\end{equation}
Because we have already fitted all parameters in the chiral Lagrangian,
our parameters in the chiral unitary model
with $SU(3)$ breaking effects are only the subtraction constants.


In the chiral Lagrangian, 
there are other $\mathcal{O}(p^2)$ terms symmetric in the $SU(3)$
flavor in addition to the above breaking terms,
if we strictly follow the ordinary chiral counting rule in powers of
the pseudoscalar meson momentum $p$ and the quark mass $m$,
where the GMOR relation fixes the ratio of $m$ and $p^{2}$.
Indeed, it is known in chiral perturbation theory that 
at $\mathcal{O}(p^2)$
the $\pi N$ scattering length is correctly obtained through
a large cancellation between the $SU(3)$ breaking term and a symmetric
term~\cite{Bernard:1993fp,Bernard:1995pa}, because
the lowest-order, {\it i.e.}
the Weinberg-Tomozawa term, already provides a sufficiently good result.
This would imply that only the inclusion of the breaking term would
be inconsistent with
the cancellation.

However, in the present work,
the symmetric terms are not taken into account for
the following reasons.
1) These terms are not responsible
for the symmetry breaking which we would like to study in this paper.
2) The purpose of the present work is to investigate
baryon resonances as dynamically generated objects.
The symmetric terms of order 
$\mathcal{O}(p^{2})$ may contain
information regarding resonances~\cite{Bernard:1995dp},
as shown for the role of the $\rho$ meson
in $\pi$-$\pi$ scattering~\cite{Ecker:1989te}.
The inclusion of some of the symmetric
terms would introduce intrinsic properties of genuine resonances
that originate from the quarks.
3) In our calculation, the $\pi N$ scattering length is 
qualitatively reproduced well without the 
$\mathcal{O}(p^2)$ symmetric terms, because the subtraction constants
in the
chiral unitary approach are adjustable parameters determined 
by the threshold
branching ratio Eq.~\eqref{eq:branch}.
Strictly speaking, 
as argued in Ref.~\citen{Oller:2000fj}, 
the subtraction constants appear as $\mathcal{O}(p^3)$ quantities
in the chiral expansion of the amplitude obtained in the
unitary approach, because they originate from the loop integral.
Therefore, they should not cancel the
quark mass terms, which are counted as $\mathcal{O}(p^2)$. 
Nevertheless, we have room to
interpret the subtraction constants as containing some of the
$\mathcal{O}(p^2)$ terms that we do not take into account
explicitly,
as the parameter fitting is carried out for the full amplitudes
obtained in the unitarity resummation at the physical
threshold, and, as we see below, the threshold ratios are
qualitatively reproduced much better than ChPT at lowest-order.
This implies that some partial contributions of the symmetric terms
are taken into account as constant values at the threshold.

In order to demonstrate the third point above,
let us introduce another set of parameters $a^{\prime}_{i}$
that originate in the $\mathcal{T}_{ij}^{-1}$ 
term in Eq.~\eqref{eq:NDamplitude},
\begin{equation}
    T^{-1}_{ij}(\sqrt{s})
    =\delta_{ij}\Bigl(\tilde{a}_i(s_0)+\frac{s-s_0}{2\pi}
    \int_{s^{+}_i}^{\infty}ds^{\prime}
    \frac{\rho_i(s^{\prime})}{(s^{\prime}-s)(s^{\prime}-s_0)}
    \Bigr)
    +a_i^{\prime}\delta_{ij}+\mathcal{T}^{-1}_{ij} \ .
    \label{eq:NDamplitude2}
\end{equation}
Here, we assume that the parameters $a^{\prime}_{i}$ 
form a diagonal matrix in
the channel space.
Note that the parameters $a^{\prime}_{i}$ are introduced as
quantities that are not related to
the regularization of the loop integral, but they should
be determined by ChPT.
Now the parameters $a_i^{\prime}$ can be related to the coefficients 
of the $\mathcal{O}(p^2)$ symmetric Lagrangian.
They are expressed as combinations of the two meson momenta
\begin{equation}
    p_1^2 \ , \quad  p_2^2 \ , \quad
    p_1\cdot p_2 \ , \quad \sigma_{\mu\nu}p_1^{\mu}p_2^{\nu} \ ,
    \label{eq:combination}
\end{equation}
with subscripts $1$ and $2$ indicating
the initial and final states, respectively.
The last term does not contribute to the $s$-wave amplitude,
and 
due to the symmetry under interchanges of 1 and 2 mesons,
the coefficients of $p_1^2$ and $p_2^2$ should be the same.
Therefore we have two independent coefficients.
It is appropriate to consider the complete set of $p^2$ terms in the
interaction kernel in order to strictly maintain consistency with
ChPT and to achieve better agreement with the amplitudes.
Once again, however, here we would like to study the $SU(3)$ breaking
effect on the excited baryons as dynamically generated objects.
In our procedure, the $SU(3)$ breaking is considered in the chiral
perturbation theory completely, but without properties of genuine
resonances.

As seen in Eq.~\eqref{eq:NDamplitude2}, the parameters $a_i^{\prime}$
can be absorbed into the subtraction constants $\tilde{a}_i$, as
$\tilde{a}_i\to \tilde{a}_i+a_i^{\prime}$.
Furthermore, $SU(3)$ symmetry reduces $\tilde{a}_i$ to a single
parameter, $\tilde{a}$.
Hence, by adjusting $\tilde{a}$,
we can use one degree of freedom of $a^{\prime}$ to fit the low
energy data.
The introduction of $a^{\prime}$ is equivalent to the
replacement
\begin{equation}
    G \to G+a^{\prime} \ .
    \label{eq:Greplace}
\end{equation}
Now, we expand the unitarized amplitude~\eqref{eq:breakresult}
in terms of the small meson momentum $p$, assuming that $a^{\prime}$
is an $\mathcal{O}(p^0)$ quantity, as 
\begin{equation}
     \begin{split}
T&=V^{(WT)}+V^{(SB)}
+(V^{(WT)}+V^{(SB)}) (G+a^{\prime}) (V^{(WT)}+V^{(SB)})
+\dotsb \\
&=\underbrace{V^{(WT)}}_{p^1}+\underbrace{V^{(SB)}
+V^{(WT)}a^{\prime}V^{(WT)}}_{p^2}
+V^{(WT)}GV^{(WT)}
+\dotsb \ .
     \end{split}
     \label{eq:breakexpand}
\end{equation}
The third term in the second line, $V^{(WT)}a^{\prime}V^{(WT)}$,
can play the role of an
interaction derived from the $p^2$ Lagrangian
and may cancel the $V^{(SB)}$
contribution to the scattering length when we choose $\tilde{a}+a^{\prime}$
such that
the low energy amplitude is reproduced.

\subsection{The $S=-1$ channel}
\begin{figure}[tbp]
    \centering
    \includegraphics[width=8.3cm,clip]{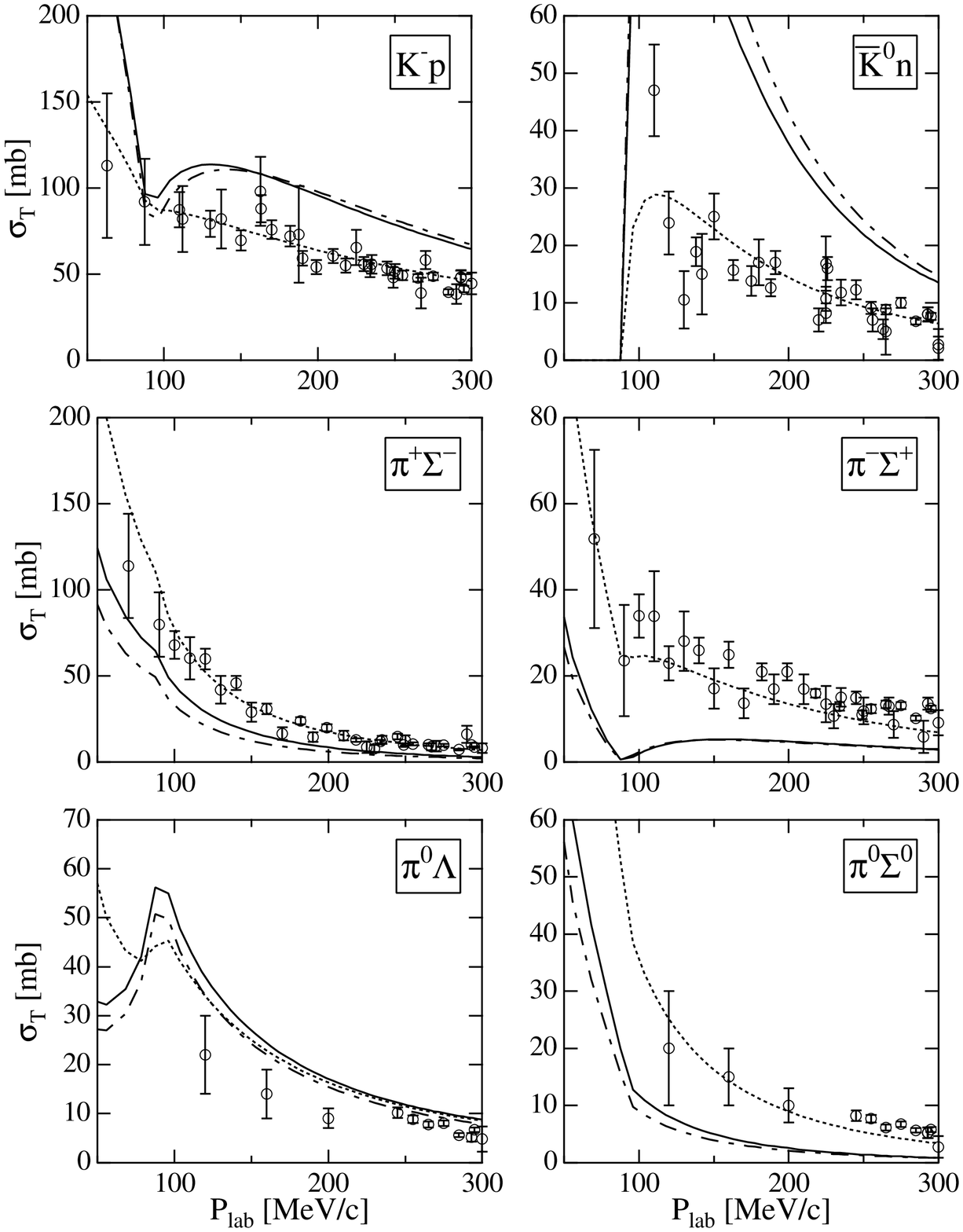}
    \caption{Total cross sections of 
    $K^{-}p$ scattering ($S=-1$)
    as functions of  $P_{\text{lab}}$,
    the three-momentum of the initial $K^{-}$
    in the laboratory frame.
    The dotted curves represent
    the results obtained with the common value $a=-1.96$,
    the dash-dotted curves represent the results obtained 
    including the $SU(3)$ breaking 
    with the common value $a=-1.59$, and
    the solid curves represent the results obtained
    including the $SU(3)$ breaking 
    and the physical $f$ with the common value $a=-1.68$.
    The open circles with error bars are experimental data
    taken from Refs.~\citen{Mast:1976pv,
    Ciborowski:1982et,Bangerter:1981px,
    Mast:1975sx,Sakitt:1965kh,
    PR131.2248,PRL8.23,PL16.89,PRL14.29,PR123.2168,PL21.349,NC16.848}.}
    \label{fig:S-1breakcross}\vspace{0.5cm}
\end{figure}

\begin{figure}[tbp]
    \centering
    \includegraphics[width=8.3cm,clip]{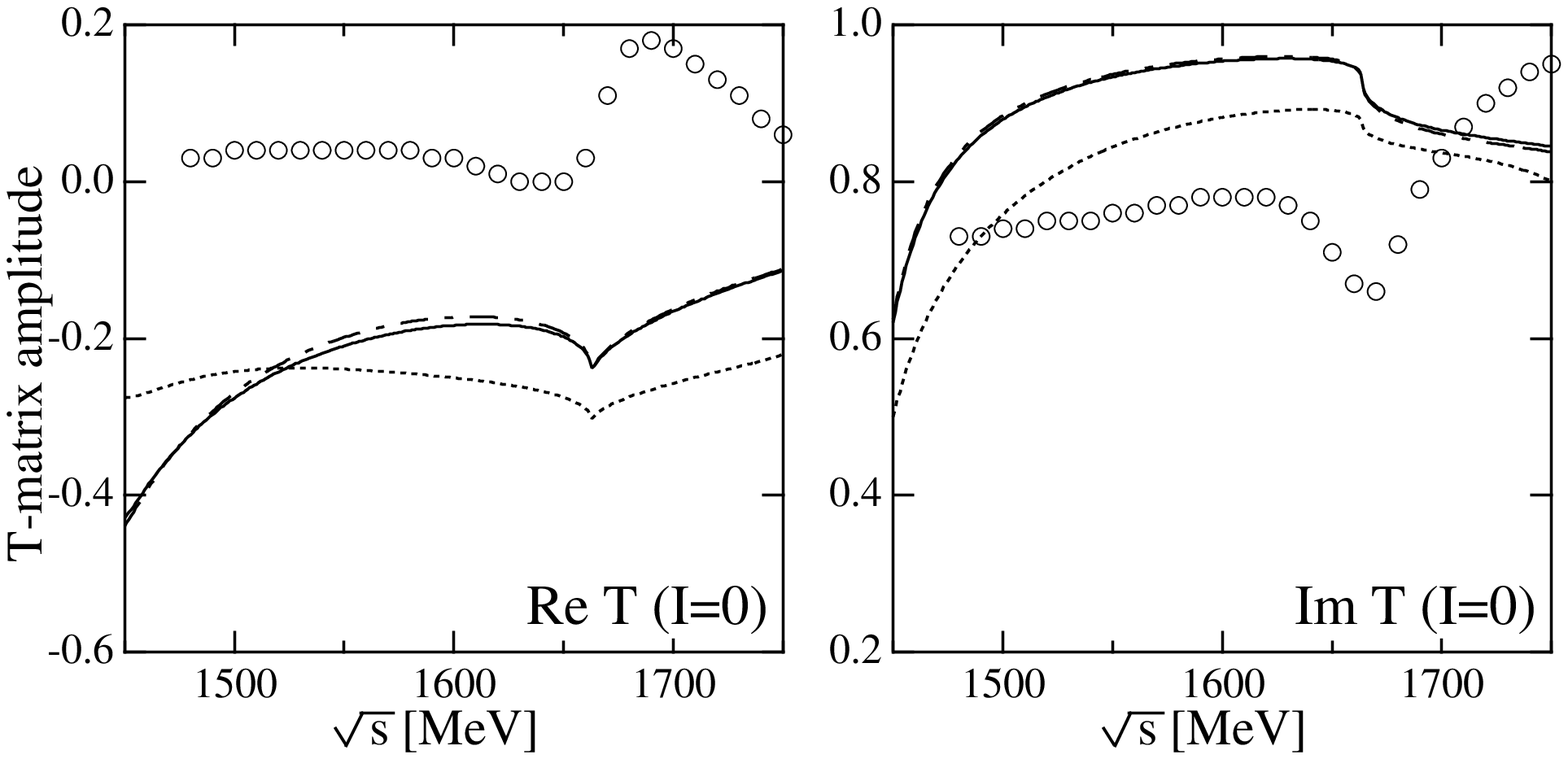}
    \caption{Real and imaginary parts of the 
    T-matrix amplitude of $\bar KN\to\bar KN$ with $I=0$.
    The dotted curves represent
    the results obtained with the common value $a=-1.96$,
    the dash-dotted curves represent
    the results obtained including the $SU(3)$ breaking 
    with the common value $a=-1.59$, and
    the solid curves represent 
    the results obtained including the $SU(3)$ breaking 
    and the physical $f$ with the common value $a=-1.68$.
    The open circles are the experimental data
    taken from Ref.~\citen{Gopal:1977gs}.}
    \label{fig:S-1breaktmat}\vspace{0.5cm}
    \includegraphics[width=8.3cm,clip]{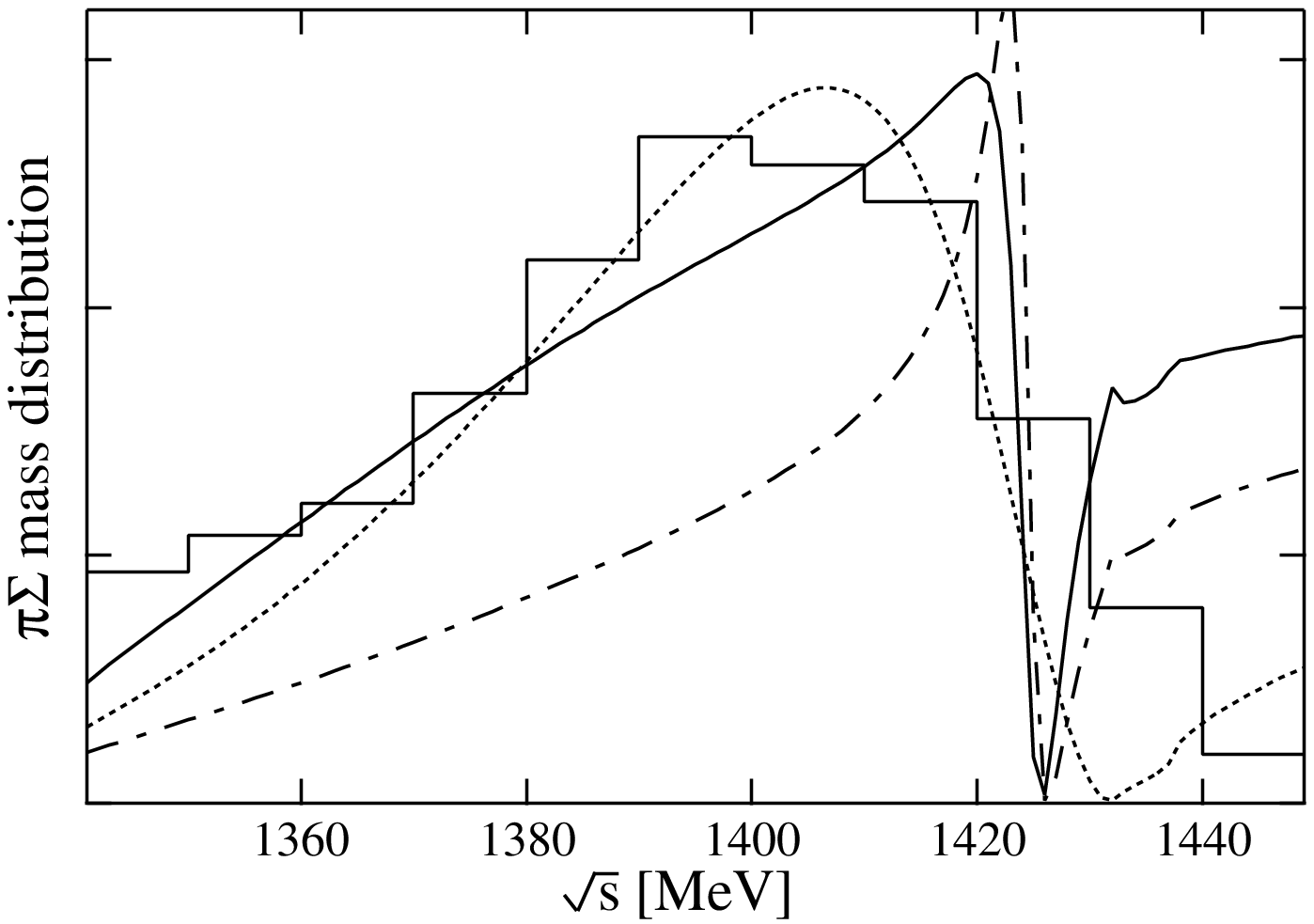}
    \caption{Mass distributions of the $\pi \Sigma$ channel
    with $I=0$.
    The dotted curve represents
    the result obtained with the common value $a=-1.96$,
    the dash-dotted curve represents
    the result obtained including the $SU(3)$ breaking 
    with the common value $a=-1.59$, and
    the solid curve represents
    the result obtained including the $SU(3)$ breaking 
    and the physical $f$ with the common value $a=-1.68$.
    The histogram represents the experimental data
    taken from Ref.~\citen{Hemingway:1985pz}.}
    \label{fig:S-1breakmdist}
\end{figure}

We follow the same procedures here as in the calculations 
without the $SU(3)$ breaking terms. 
First, we determine the common subtraction constant $a$
from the threshold branching ratios \eqref{eq:branch}.
The optimal value
is found to be $a=-1.59$. 
With this value, the total cross sections of
the $K^- p$ scattering, the $\pi \Sigma$ mass distribution, and the
scattering amplitude of $\bar K N \to \bar K N$ with $I=0$ are plotted in
Figs.~\ref{fig:S-1breakcross}, \ref{fig:S-1breaktmat}
and \ref{fig:S-1breakmdist}
by the dash-dotted curves.
As seen in Fig.~\ref{fig:S-1breakcross}, for all the total cross sections,
the inclusion of the $SU(3)$ breaking terms with the common value $a$
causes the agreement with data to become worse,
although the threshold branching ratios are produced much better
than in the previous works, as seen in Table \ref{tbl:branch}.

In the $\pi \Sigma$ mass distribution shown in 
Fig.~\ref{fig:S-1breakmdist} (the dash-dotted curve), 
a sharp peak is seen,
in obvious contradiction with 
the observed spectrum.
This means that the important 
resonance structure of $\Lambda(1405)$ has been lost.
However, we find two poles of the T-matrix amplitude at 
$z_1=1424-1.6i$ and $z_2=1389-135i$
in the second Riemann sheet. 
It is reported that there are two poles
in the T-matrix amplitude around the energy region of $\Lambda(1405)$
in Refs.~\citen{Fink:1990uk,Oller:2000fj,Jido:2002yz,Garcia-Recio:2002td,
Nam:2003ch,Garcia-Recio:2003ks}.
A detailed study of the two poles for $\Lambda(1405)$ has recently been
done from the viewpoint of the $SU(3)$ flavor symmetry
in Ref.~\citen{Jido:2003cb},
and it has also been argued 
in the case of reaction processes~\cite{Hyodo:2003jw,Hyodo:2004vt}.
The inclusion of the $SU(3)$ breaking terms does not
change this conclusion,
although the positions of the poles change.

We also calculate the total cross sections and the $\pi \Sigma$ 
mass distribution with  the physical values of the meson decay
constants,  $f_{\pi}=93\text{ MeV},\;
f_{K}=1.22f_{\pi},\;f_{\eta}=1.3f_{\pi}$. 
The calculated results are represented in Figs.~\ref{fig:S-1breakcross},
\ref{fig:S-1breaktmat} and \ref{fig:S-1breakmdist}
by the solid curves. 
The optimal value of the subtraction constants is $a=-1.68$,
and with this value,
the threshold branching ratios are reproduced as
$\gamma= 2.35$, $R_c  = 0.626$ and $R_n = 0.172$.
The $SU(3)$ breaking effect on the meson 
decay constants is not so large in the total cross sections,
as seen in the figures. 
However, the shape of the peak seen in the $\pi \Sigma$ mass distribution
becomes wider than that in the calculation
with the averaged meson decay constant.

Indeed, we again find two poles in the scattering amplitudes
at $z_1^{\prime}=1424-2.6i$ and $z_2^{\prime}=1363-87i$
in the second Riemann sheet.
Compared with the poles $z_1$ and $z_2$
obtained in the above calculation, the position of the pole $z_2^{\prime}$ 
moves to the lower energy side and approaches the real axis.
The reason why the position of $z_2^{\prime}$ changes
can be understood as follows.
Because $z_2$ has a large imaginary part, which implies a large width,
and only the $\pi\Sigma$ channel is open in this energy region,
the resonance represented by the pole $z_2$ has a strong coupling
to the $\pi\Sigma$ channel.
This fact implies that the position of the pole $z_2$ is
sensitive to the $\pi\Sigma$ interaction.
In the present calculation, the pion decay constant (93 MeV)
is smaller than the averaged value (106.95 MeV)
used in the above calculation,
so that the attractive interaction of $\pi\Sigma$ becomes stronger.
This shifts the position of the pole $z_2$ to the lower energy side.
Simultaneously, 
this suppresses the phase space of the decay of the resonance 
to the $\pi\Sigma$ channel,
and hence, the position of the pole approaches the real axis.


\subsection{The $S=0$ channel}
Here we present
calculations in the $S=0$ channel with the $SU(3)$ breaking terms.
With a common value $a\sim -1.5$,
with which the threshold properties are reproduced
well in the $S=-1$ channel, we still obtain a large contribution
in the $S_{11}$ $\pi N$ scattering amplitude 
at low energy, as in the calculation without the 
$SU(3)$ breaking effects. 
From this analysis, it is found that the low energy behavior
of the $\pi N$ scattering cannot be reproduced 
as long as we use the common value $a \sim -2$,
even if we introduce the $SU(3)$ breaking effects.

\begin{figure}[tbp]
    \centering
    \includegraphics[width=8.3cm,clip]{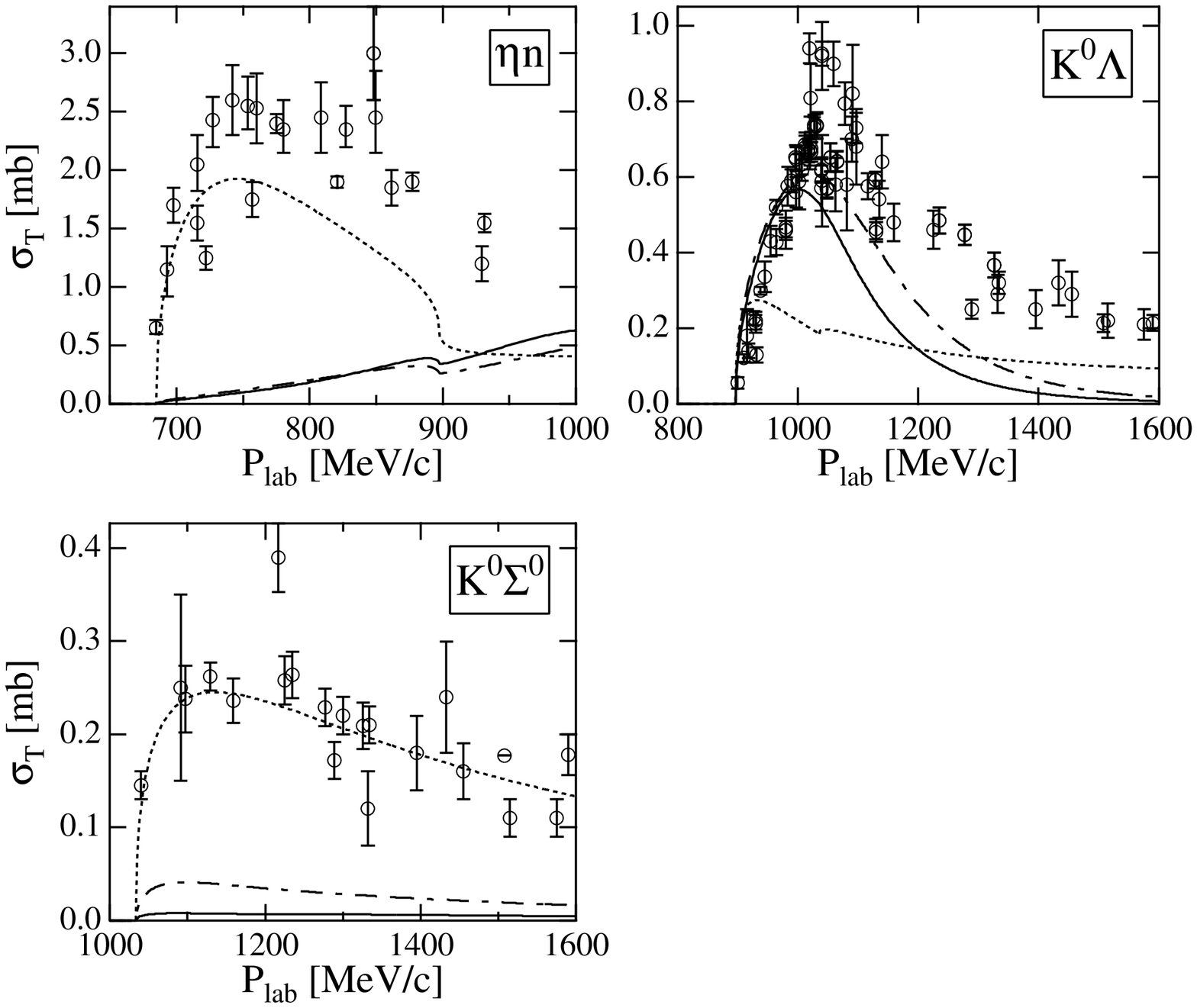}
    \caption{Total cross sections of 
    $\pi^{-} p$ scattering ($S=0$)
    as functions of  $P_{\text{lab}}$,
    the three-momentum of the initial $\pi^{-}$
    in the laboratory frame.
    The dotted curve represent
      the results obtained with the common value $a=0.53$,
    the dash-dotted curves represent the results
     obtained including the $SU(3)$ breaking interaction
      with the common value $a=1.33$, and
    the solid curves represent the results obtained
    including the $SU(3)$ breaking 
    and the physical $f$ with the common value $a=2.24$.
    The open circles with error bars are experimental data
    taken from Refs.~\citen{Batinic:1995kr,Hart:1980jx,Saxon:1980xu,
    Baker:1978bb,Baker:1978qm,
    Nelson:1973py,Thomas:1973uh,
    Jones:1971zm,Binford:1969ts,VanDyck:1969ay,
    PRD11.1,PR155.1455,PR133.B457,PR132.1778,
    PR109.1358,PRL8.332,PRL3.394,
    NC53A.745,NC42A.606,NC10.468}.}
    \label{fig:S0breakcross} \vspace{0.5cm}
    \centering
    \includegraphics[width=8.3cm,clip]{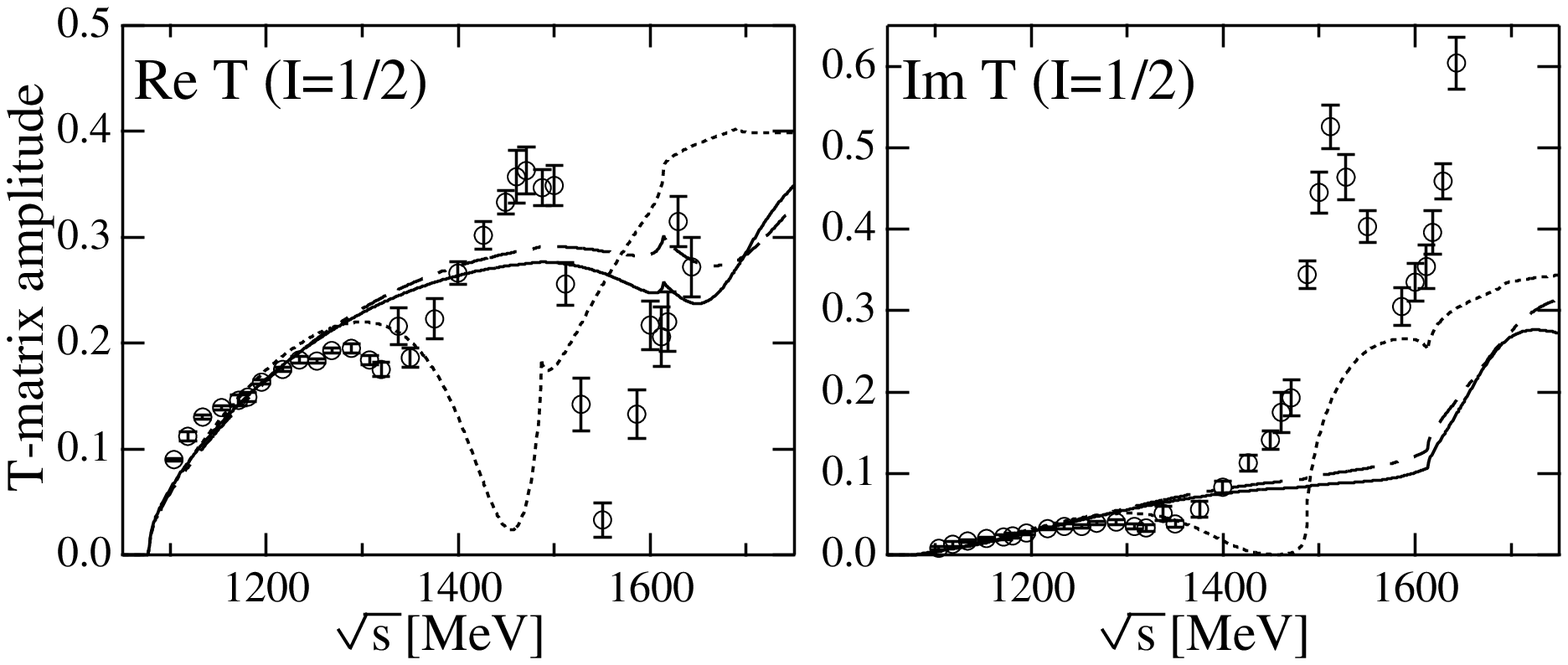}
    \caption{Real and imaginary parts of the $S_{11}$
    T-matrix amplitudes of
    $\pi N\to \pi N$.
    The dotted curves represent
      the results obtained with the common value $a=0.53$,
    the dash-dotted curves represent the results
     obtained including the $SU(3)$ breaking interaction
      with the common value $a=1.33$, and
    the solid curves represent the results obtained
    including the $SU(3)$ breaking 
    and the physical $f$ with the common value $a=2.24$.
    The open circles with error bars are experimental data
    taken from Ref.~\citen{CNS}.}
    \label{fig:S0breaktmat}
\end{figure}

In order to search for the optimal value of the common
subtraction constant
within the $S=0$ channel, we carried out a
fitting of the T-matrix elements
in the $\pi N$ $S_{11}$ channel
in the low energy region up to 1400 MeV.
We find $a=1.33$.
The results including the $SU(3)$ breaking effects with
$a=1.33$ are represented as dash-dotted curves 
in Figs.~\ref{fig:S0breakcross} and \ref{fig:S0breaktmat}.
As seen in Fig.  \ref{fig:S0breaktmat}, the fitting is accurate 
up to $\sqrt s \sim 1400$ MeV, while, however, the resonance 
structure does not appear near energies of $N(1535)$.

Finally, we present the calculations
with the physical values of the meson decay constants
in Figs.~\ref{fig:S0breakcross} and \ref{fig:S0breaktmat}
(solid curves). The optimal value of the
common subtraction constant is found to be $a=2.24$.
The results with the 
physical meson decay constants and $a=2.24$
are very similar to the results of the calculation
with the averaged value of the decay constants and $a=1.33$.
In this sense, the $SU(3)$ breaking effect
of the meson decay constant $f$ is absorbed 
into the change of the common subtraction constant $a$.  

In closing this section,
we conclude that even if we introduce the $SU(3)$ breaking
effects at the Lagrangian level,  the $SU(3)$ breaking
in the channel-dependent subtraction
constants $a_i$ cannot be absorbed into
the $SU(3)$ breaking effects in the 
fundamental interactions in both the $S=-1$ and $S=0$ channels.

\section{Summary and discussion}\label{sec:discussion}
In this work, first we attempted to use
a single common subtraction constant
in order to describe meson-baryon scattering
and baryon resonance in a unified way.
In the $S=-1$ channel, $a \sim -2$ is fixed from the threshold 
branching ratios of the $K^- p$ scattering.
With this parameter values, 
the total cross sections of the $K^- p$ scattering are 
reproduced well, as well as the mass distribution for $\Lambda(1405)$.
However, in this case 
the $\Lambda(1670)$ resonance cannot be reproduced. 
The subtraction constant $a\sim -2$ corresponds to
$\Lambda = 630$ MeV in the three-momentum cut-off regularization 
of the meson-baryon loop integral~\cite{Oller:2000fj}.
This value is consistent with that often
used in single nucleon processes~\cite{Bennhold:1991kj}.
The elementary interaction of the $\bar K N$ system 
is sufficiently attractive,
and a resummation of the coupled channel interactions 
causes the $\Lambda(1405)$ resonance to appear
at the correct position,
by imposing the unitarity condition
and by using the natural value for the cut-off parameter.
Hence, the wave function of $\Lambda(1405)$ is largely
dominated by the $\bar K N$ component.

On the other hand, in the $S=0$ channel,
if one uses the natural value for the subtraction constant,
as in the $S=-1$ channel,
the attraction of the meson-baryon interaction becomes so 
strong that an unexpected resonance is generated near
$\sqrt s  \sim 1250$ MeV. 
Therefore, a repulsive component is
necessary to reproduce the observed $\pi N$ scattering.
The fitted subtraction constant using the low energy 
$\pi N$ scattering amplitude is $a\sim 0.5$. 
With this value, however, the $N(1535)$ resonance
is not generated,
while the agreement among the cross sections
of $\pi^{-}p \to \eta n$ is rather good,
due to the threshold effects.

The unitarized amplitudes are very sensitive
to the attractive component of the interaction.
The interaction terms of the ChPT alone do not explain
all scattering amplitudes simultaneously.
Rather, they must be complemented
by subtraction constants in the chiral unitary model.
For small $a$, the interaction becomes more attractive,
and for large $a$, less attractive.
For $S=0$, we need to choose $a\sim 0.5$
in order to suppress the attraction
from the $\pi N$ interaction, in contrast to 
the situation for the natural value $a\sim -2$ in the $S=-1$ channel.
Therefore, it is not possible to reproduce both the
$\Lambda(1405)$ resonance properties and the low energy $\pi N$ 
scattering with a common subtraction constant
within the present framework.

Generally speaking, the chiral unitary approach is  
a powerful phenomenological method.
It can reproduce cross sections and generate
$s$ wave resonances dynamically,
once the subtraction constants are determined appropriately,
using experimental data.
However, it is not straightforward
to apply the method to channels
for which there are not sufficient experimental data,
because they are needed to determine the subtraction constants,
unless we employ a channel-independent renormalization scheme,
as in Refs.~\citen{Lutz:2000us,Lutz:2001yb}
and \citen{Garcia-Recio:2003ks}.

Next, we introduced the flavor $SU(3)$ breaking Lagrangian,
with the hope that the channel dependence in the subtraction constants 
would be absorbed into the coefficients in the chiral Lagrangian.
These coefficients can be determined from other observables, 
and hence they are more controllable 
than the subtraction constants,
which have to be fitted to the experimental 
data. However, the channel dependence of the subtraction constants
in each strangeness channel cannot be replaced 
by the $SU(3)$ breaking Lagrangian,
although we have exhausted possible breaking sources
up to order $\mathcal{O}(m_{q})$.

Therefore, in the present framework, in which
the Weinberg-Tomozawa term and
symmetry breaking terms are taken into account, 
a suitable choice of the channel-dependent
subtraction constants is essential.
Theoretically, it would be very important to obtain
a microscopic explanation of the origin of the channel-dependent
subtraction constants.
One possibility is to consider quark degrees of freedom, which can
generate genuine resonance states.
Another possibility to solve this problem is to employ 
interaction terms
up to order $p^3$ with the channel-independent
renormalization scheme~\cite{Lutz:2001yb}.
Further investigations should be carried out
in order to better understand the nature of baryon resonances.

\section*{Acknowledgements}
We would like to thank Profs.\ E.\ Oset, H.\ -Ch.\ Kim 
and W.\ Weise for useful
discussions.

\appendix

\section{Coefficients of the $SU(3)$ Breaking
Interaction}\label{sec:coefficients}

\begin{table}[tbp]
    \caption{Channels of meson-baryon scattering.
    In this work, we carried out calculation for the channels in
    $(S=-1,Q=0)$ and $(S=0,Q=0)$.}
    \label{tab:classification}
\begin{tabular}{|cc|cc|l|}
    \hline
    $Y$ & $S$ & $I_3$ & $Q$  & channels  \\
    \hline
    $-2$ & $-3$ & $1$ & $0$ & $\bar{K}^{0}\Xi^{0}$   \\
    & & $0$ & $-1$ & $K^{-}\Xi^{0},\;\;\bar{K}^{0}\Xi^{-}$   \\
    & & $-1$ & $-2$ & $K^{-}\Xi^{-}$   \\
    \hline
    $-1$ & $-2$ & $\frac{3}{2}$ & $1$ 
	& $\pi^{+}\Xi^{0},\;\;\bar{K}^{0}\Sigma^{+}$ \\
    & & $\frac{1}{2}$ & $0$ & $\pi^{0}\Xi^{0},\;\;\pi^{+}\Xi^{-},\;\;
	\eta\Xi^{0},\;\;\bar{K}^{0}\Lambda,\;\;\bar{K}^{0}\Sigma^{0},
	\;\;K^{-}\Sigma^{+}$ \\
    & & $-\frac{1}{2}$ & $-1$ & $\pi^{0}\Xi^{-},\;\;\pi^{-}\Xi^{0},\;\;
	\eta\Xi^{-},\;\;K^{-}\Lambda,\;\;K^{-}\Sigma^{0},\;\;\bar{K}^{0}\Sigma^{-}$  \\
    & & $-\frac{3}{2}$ & $-2$ & $\pi^{-}\Xi^{-},\;\;K^{-}\Sigma^{-}$ \\
    \hline
    $0$ & $-1$ & $2$ & $2$ & $\pi^{+}\Sigma^{+}$  \\
    & & $1$ & $1$ & $\bar{K}^{0}p,\;\;\pi^{0}\Sigma^{+},\;\;\pi^{+}\Sigma^{0},\;\;
	\pi^{+}\Lambda,\;\;\eta\Sigma^{+},\;\;K^{+}\Xi^{0}$   \\
    & & $0$ & $0$ & $K^{-}p,\;\;\bar{K}^{0}n,\;\;\pi^{0}\Lambda,
    \;\;\pi^{0}\Sigma^{0},\;\;
	\eta\Lambda,\;\;\eta\Sigma^{0},\;\;\pi^{+}\Sigma^{-},\;\;
	\pi^{-}\Sigma^{+},\;\;K^{+}\Xi^{-},\;\;K^{0}\Xi^{0}$   \\
    & & $-1$ & $-1$ & $K^{-}n,\;\;\pi^{0}\Sigma^{-},\;\;\pi^{-}\Sigma^{0},\;\;
	\pi^{-}\Lambda,\;\;\eta\Sigma^{-},\;\;K^{0}\Xi^{-}$ \\
    & & $-2$ & $-2$ & $\pi^{-}\Sigma^{-}$   \\
    \hline
    $1$ & 0 & $\frac{3}{2}$ & $2$ & $\pi^{+}p,\;\;K^{+}\Sigma^{+}$  \\
    & & $\frac{1}{2}$ & $1$ & $\pi^{0}p,\;\;\pi^{+}n,\;\;\eta p,\;\;
	K^{+}\Lambda,\;\;K^{+}\Sigma^{0},\;\;K^{0}\Sigma^{+}$  \\
    & & $-\frac{1}{2}$ & $0$ & $\pi^{0}n,\;\;\pi^{-}p,\;\;\eta n,\;\;
	K^{0}\Lambda,\;\;K^{0}\Sigma^{0},\;\;K^{+}\Sigma^{-}$ \\
    & & $-\frac{3}{2}$ & $-1$ & $\pi^{-}n,\;\;K^{0}\Sigma^{-}$ \\
    \hline
    $2$ & $1$ & $1$ & $2$ & $K^{+}p$   \\
    & & $0$ & $1$ & $K^{+}n,\;\;K^{0}p$  \\
    & & $-1$ & $0$ & $K^{0}n$  \\ \hline
\end{tabular}    
\end{table}%

Here, we derive the coefficients of the
flavor $SU(3)$ breaking terms
in the meson-baryon interactions.
The corresponding Lagrangian is given by
\begin{align}
     \mathcal{L}_{SB}^{(2)}
     =&\frac{Z_0}{4f^{2}}\Tr\Bigl(
     d_m\bar{B}\bigl\{(2\Phi\bm{m}\Phi
     +\Phi^{2}\bm{m}
     +\bm{m}\Phi^{2}),B\bigr\}
     +f_m\bar{B}\bigl[(2\Phi\bm{m}\Phi
     +\Phi^{2}\bm{m}
     +\bm{m}\Phi^{2}),B\bigr]
     \Bigr)\notag \\*
     &+\frac{Z_1}{f^{2}}
     \Tr(\bar{B}B)\Tr(\bm{m}\Phi^{2}) \ .
     \label{eq:SBLagApp}
\end{align}
From this Lagrangian,
the basic interaction at tree level
is given by
\begin{align}
     V^{(SB)}_{ij}=&-\frac{1}{f^{2}}
     \Bigl[
     Z_0\Bigl( (A^{d}_{ij}d_m
     +A^{f}_{ij}f_m )\hat{m}
     +(B^{d}_{ij}d_m
     +B^{f}_{ij}f_m)m_s
     \Bigr)\notag \\*
     &+Z_1\delta_{ij}D^{Z_1}_{i}\Bigr]
     \sqrt{\frac{E_i+M_i}{2M_i}}
     \sqrt{\frac{E_j+M_j}{2M_j}},
\end{align}
where the coefficients $A$, $B$ and $D$ are the numbers in matrix
form,
and the indices $(i,j)$ denote the channels
of the meson-baryon scattering,
as shown in Table~\ref{tab:classification}.

These channels are specified
by two quantum numbers,
the hypercharge, $Y$, and the third component
of isospin $I_3$, or equivalently
the strangeness, $S$,
and the electric charge, $Q$,
through the Gell-Mann--Nakano--Nishijima relation~\cite{PR92.833,PTP10.581}
\begin{equation}
    Q=T_3+\frac{Y}{2} \ , \quad
    S=Y-B \ ,
\end{equation}
where the baryon number is $B=1$
for the meson-baryon scattering.

The coefficient $D^{Z_1}_{i}$ is specified only by the meson
in channel $i$, independently of the baryons,
because $\Tr(\bar{B}B)$ in the last term of Eq.~\eqref{eq:SBLagApp}
gives a common contribution
to all baryons.
Also, there is no off-diagonal component
when the isospin symmetry is assumed.
\begin{table}[tbp]   
    \centering
    \caption{Form of $D^{Z_1}_{i}$.}
    \label{tab:Bz1}
    \begin{tabular}{|c|ccc|}\hline
	meson & $\pi$ & $K,\bar{K}$  & $\eta$   \\ \hline
	$D^{Z_1}_{i}$ & $2\hat{m}$ & $\hat{m}+m_s$
	& $\frac{2}{3}(\hat{m}+2m_s)$ \\ \hline
    \end{tabular}
\end{table}%
The explicit form of $D^{Z_1}_{i}$ is presented in Table~\ref{tab:Bz1}.
The values of the coefficients $A$ and $B$
are given in subsequent tables, as follows:
\begin{itemize}
    \item  Table~\ref{tab:ABS1} ($S=1,Q=1$)

    \item  Table~\ref{tab:ABS-3} ($S=-3,Q=-1$)
    
    \item  Tables~\ref{tab:AS0} and \ref{tab:BS0} ($S=0,Q=0$)

    \item  Tables~\ref{tab:AS-2} and \ref{tab:BS-2} ($S=-2,Q=-1$)

    \item  Tables~\ref{tab:AdS-1},\ref{tab:AfS-1},
	    \ref{tab:BdS-1} and \ref{tab:BfS-1} ($S=-1,Q=0$).
\end{itemize}
From these tables, the coefficients $A$ and $B$ 
for all the channels can be derived,
using symmetry relations.

First, channels with equal $S$ and unequal $Q$ are
related through the SU(2) Clebsch-Gordan coefficients
due to the isospin symmetry.
This is the relation among the channels
in the block separated by the horizontal lines
in Table~\ref{tab:classification}.
Second, the coefficients of the sector ($Y,I_3$)
are related with those of ($-Y,-I_3$).
Let us consider the channels $(i,j)$
and $(i^{\prime},j^{\prime})$
in the sectors ($Y,I_3$) and ($-Y,-I_3$),
respectively, as shown in Table~\ref{tab:qnumber}.
\begin{table}[tbp]
    \centering
    \caption{Quantum numbers for the channels
    $i,j,i^{\prime}$ and $j^{\prime}$}
    \label{tab:qnumber}
\begin{tabular}{|c|ccc|ccc|}\hline
    channel & \multicolumn{3}{|c|}{hypercharge}  &
     \multicolumn{3}{c|}{third component of isospin} \\
    \hline
     & meson  & baryon  & total
    & meson  & baryon  & total  \\
    \hline
    $i$ & $y_i$ & $Y-y_i$ & $Y$ &
    $i_{3i}$ & $I_3-i_{3i}$ & $I_3$ \\
    \hline
    $j$ & $y_j$ & $Y-y_j$ & $Y$ &
    $i_{3j}$ & $I_3-i_{3j}$ & $I_3$ \\
    \hline
    $i^{\prime}$ & $-y_i$ & $-Y+y_i$ & $-Y$ &
    $-i_{3i}$ & $-I_3+i_{3i}$ & $-I_3$ \\
    \hline
    $j^{\prime}$ & $-y_j$ & $-Y+y_j$ & $-Y$ &
    $-i_{3j}$ & $-I_3+i_{3j}$ & $-I_3$ \\
    \hline
\end{tabular}  
\end{table}%
Then, the coefficients of the sector $(-Y,-I_3)$ are given by
\begin{equation}
    \begin{split}
    A^{d}_{i^{\prime}j^{\prime}}(-Y,-I_3)&=
    A^{d}_{ij}(Y,I_3), \quad
    A^{f}_{i^{\prime}j^{\prime}}(-Y,-I_3)=
    -A^{f}_{ij}(Y,I_3), \\
    B^{d}_{i^{\prime}j^{\prime}}(-Y,-I_3)&=
    B^{d}_{ij}(Y,I_3), \quad
    B^{f}_{i^{\prime}j^{\prime}}(-Y,-I_3)=
    -B^{f}_{ij}(Y,I_3).        
    \end{split}
    \label{eq:relation}
\end{equation}
Comparing Table~\ref{tab:ABS1} ($Y=2,I_3=0$) and
Table~\ref{tab:ABS-3} ($Y=-2,I_3=0$),
we find that the relation \eqref{eq:relation}
is satisfied.
Also, using the relation \eqref{eq:relation},
the coefficients of the sector $(S=-2,Q=0)$
are obtained from the tables
of the sector $(S=0,Q=0)$.
For example, if we specify 
$(i,j)$ to be $(\pi^{0}n,K^{0}\Lambda)$,
the corresponding $(i^{\prime},j^{\prime})$ 
is $(\pi^{0}\Xi^{0},\bar{K}^{0}\Lambda)$.
The coefficients for $(i^{\prime},j^{\prime})$
are obtained as
$A_{i^{\prime}j^{\prime}}^{d}=\sqrt{3}/8$,
$A_{i^{\prime}j^{\prime}}^{f}=-3\sqrt{3}/8$,
$B_{i^{\prime}j^{\prime}}^{d}=1/(8\sqrt{3})$ and 
$B_{i^{\prime}j^{\prime}}^{f}=-\sqrt{3}/8$.
In this way, we can derive all the coefficients
that are not shown in the tables.

\begin{table}[tbp]
    \centering
    \caption{$A^{d}_{ij},A^{f}_{ij},B^{d}_{ij}$
     and $B^{f}_{ij}(S=1,Q=1)$}
    \label{tab:ABS1}
\begin{tabular}{|c|cc|cc|cc|cc|}\hline
    & \multicolumn{2}{|c|}{$A^{d}_{ij}$}
    & \multicolumn{2}{c|}{$A^{f}_{ij}$}
    & \multicolumn{2}{c|}{$B^{d}_{ij}$}
    & \multicolumn{2}{c|}{$B^{f}_{ij}$}\\ \hline
    & $K^{+}n$ & $K^{0}p$  & $K^{+}n$ & $K^{0}p$
    & $K^{+}n$ & $K^{0}p$  & $K^{+}n$ & $K^{0}p$ \\ \hline
$K^{+}n$   & $\frac{1}{2}$ & $\frac{1}{2}$
	   & $-\frac{1}{2}$ & $\frac{1}{2}$
       & $\frac{1}{2}$ & $\frac{1}{2}$
	   & $-\frac{1}{2}$ & $\frac{1}{2}$ \\
$K^{0}p$   & & $\frac{1}{2}$
	   & & $-\frac{1}{2}$
       & & $\frac{1}{2}$
	   & & $-\frac{1}{2}$ \\ \hline
\end{tabular}
    \caption{$A^{d}_{ij},A^{f}_{ij},B^{d}_{ij}$
     and $B^{f}_{ij}(S=-3,Q=-1)$}
    \label{tab:ABS-3}
\begin{tabular}{|c|cc|cc|cc|cc|}\hline
    & \multicolumn{2}{|c|}{$A^{d}_{ij}$}
    & \multicolumn{2}{c|}{$A^{f}_{ij}$}
    & \multicolumn{2}{c|}{$B^{d}_{ij}$}
    & \multicolumn{2}{c|}{$B^{f}_{ij}$}\\ \hline
    & $K^{-}\Xi^{0}$ & $\bar{K}^{0}\Xi^{-}$
    & $K^{-}\Xi^{0}$ & $\bar{K}^{0}\Xi^{-}$
    & $K^{-}\Xi^{0}$ & $\bar{K}^{0}\Xi^{-}$
    & $K^{-}\Xi^{0}$ & $\bar{K}^{0}\Xi^{-}$ \\ \hline
$\bar{K}^{0}\Xi^{-}$   & $\frac{1}{2}$ & $\frac{1}{2}$
	   & $\frac{1}{2}$ & $-\frac{1}{2}$
       & $\frac{1}{2}$ & $\frac{1}{2}$
	   & $\frac{1}{2}$ & $-\frac{1}{2}$ \\
$K^{-}\Xi^{0}$   & & $\frac{1}{2}$
	   & & $\frac{1}{2}$
       & & $\frac{1}{2}$
	   & & $\frac{1}{2}$ \\
	   \hline
\end{tabular}    
\end{table}%

\begin{table}[tbp]
    \centering
    \caption{$A^{d}_{ij}$ and $A^{f}_{ij}(S=0,Q=0)$}
    \label{tab:AS0}
\begin{tabular}{|c|cccccc|cccccc|}\hline
    & \multicolumn{6}{c|}{$A^{d}_{ij}$}
    & \multicolumn{6}{c|}{$A^{f}_{ij}$}  \\ \hline
    & $\pi^{0}n$ & $\pi^{-}p$  & $\eta n$
    & $K^{0}\Lambda$ & $K^{0}\Sigma^{0}$ & $K^{+}\Sigma^{-}$
   & $\pi^{0}n$ & $\pi^{-}p$  & $\eta n$
    & $K^{0}\Lambda$ & $K^{0}\Sigma^{0}$ & $K^{+}\Sigma^{-}$ \\ \hline
$\pi^{0}n$   & $1$ & $0$ & $-\frac{1}{\sqrt{3}}$
	     & $\frac{\sqrt{3}}{8}$ & $\frac{3}{8}$
	 & $\frac{3}{4\sqrt{2}}$
	 & $1$ & $0$ & $-\frac{1}{\sqrt{3}}$
	     & $\frac{3\sqrt{3}}{8}$ & $-\frac{3}{8}$
	 & $-\frac{3}{4\sqrt{2}}$ \\
$\pi^{-}p$   & & $1$ & $\sqrt{\frac{2}{3}}$ & $-\frac{\sqrt{6}}{8}$
	     & $\frac{3}{4\sqrt{2}}$ & $0$
	 & & $1$ & $\sqrt{\frac{2}{3}}$
	 & $-\frac{3\sqrt{6}}{8}$
	     & $-\frac{3}{4\sqrt{2}}$ & $0$ \\
$\eta n$   & & & $\frac{1}{3}$ & $-\frac{1}{24} $
       & $-\frac{1}{8\sqrt{3}}$ & $\frac{1}{4\sqrt{6}}$
       & & & $\frac{1}{3}$ & $-\frac{1}{8} $
       & $\frac{1}{8\sqrt{3}}$ & $-\frac{1}{4\sqrt{6}}$ \\
$K^{0}\Lambda$& & & & $\frac{5}{6}$
	   & $-\frac{1}{2\sqrt{3}}$ & $\frac{1}{\sqrt{6}}$
       & & & & $0$
	   & $0$ & $0$ \\
$K^{0}\Sigma^{0}$ & & & & & $\frac{1}{2}$ & $0$
	    & & & & & $0$ & $\frac{1}{\sqrt{2}}$\\
$K^{+}\Sigma^{-}$ & & & & & & $\frac{1}{2}$
	     & & & & & & $-\frac{1}{2}$\\
	     \hline
\end{tabular}
    \caption{$B^{d}_{ij}$ and $B^{f}_{ij}(S=0,Q=0)$}
    \label{tab:BS0}
\begin{tabular}{|c|cccccc|cccccc|}\hline
    & \multicolumn{6}{|c|}{$B^{d}_{ij}$}
    & \multicolumn{6}{c|}{$B^{f}_{ij}$}  \\ \hline
    & $\pi^{0}n$ & $\pi^{-}p$  & $\eta n$
    & $K^{0}\Lambda$ & $K^{0}\Sigma^{0}$ & $K^{+}\Sigma^{-}$
    & $\pi^{0}n$ & $\pi^{-}p$  & $\eta n$
    & $K^{0}\Lambda$ & $K^{0}\Sigma^{0}$ & $K^{+}\Sigma^{-}$ \\ \hline
$\pi^{0}n$ & $0$ & $0$ & $0$ & $\frac{1}{8\sqrt{3}}$
	   & $\frac{1}{8}$ & $\frac{1}{4\sqrt{2}}$
	   & $0$ & $0$ & $0$ & $\frac{\sqrt{3}}{8}$
	   & $-\frac{1}{8}$ & $-\frac{1}{4\sqrt{2}}$ \\
$\pi^{-}p$ & & $0$ & $0$ & $-\frac{1}{4\sqrt{6}}$
	   & $\frac{1}{4\sqrt{2}}$ & $0$
	   & & $0$ & $0$ & $-\frac{\sqrt{6}}{8}$
	   & $-\frac{1}{4\sqrt{2}}$ & $0$ \\
$\eta n$   & & & $\frac{4}{3}$ & $\frac{5}{24} $
       & $\frac{5}{8\sqrt{3}}$ & $-\frac{5}{4\sqrt{6}}$
	   & & & $-\frac{4}{3}$ & $\frac{5}{8} $
       & $-\frac{5}{8\sqrt{3}}$ & $\frac{5}{4\sqrt{6}}$ \\
$K^{0}\Lambda$& & & & $\frac{5}{6}$
	   & $-\frac{1}{2\sqrt{3}}$ & $\frac{1}{\sqrt{6}}$
	   & & & & $0$
	   & $0$ & $0$ \\
$K^{0}\Sigma^{0}$ & & & & & $\frac{1}{2}$ & $0$
	    & & & & & $0$ & $\frac{1}{\sqrt{2}}$  \\
$K^{+}\Sigma^{-}$ & & & & & & $\frac{1}{2}$
	     & & & & & & $-\frac{1}{2}$  \\
	     \hline
\end{tabular}
\end{table}
\begin{table}[tbp]
    \centering
    \caption{$A^{d}_{ij}$ and $A^{f}_{ij}(S=-2,Q=-1)$}
    \label{tab:AS-2}
\begin{tabular}{|c|cccccc|cccccc|}\hline
    & \multicolumn{6}{|c|}{$A^{d}_{ij}$}
    & \multicolumn{6}{c|}{$A^{f}_{ij}$}  \\ \hline
    & $\pi^{0}\Xi^{-}$ & $\pi^{-}\Xi^{0}$ & $\eta\Xi^{-}$
    & $K^{-}\Lambda$ & $K^{-}\Sigma^{0}$
    & $\bar{K}^{0}\Sigma^{-}$
    & $\pi^{0}\Xi^{-}$ & $\pi^{-}\Xi^{0}$ & $\eta\Xi^{-}$
    & $K^{-}\Lambda$ & $K^{-}\Sigma^{0}$
    & $\bar{K}^{0}\Sigma^{-}$   \\ \hline
$\pi^{0}\Xi^{-}$& $1$ & $0$ & $\frac{1}{\sqrt{3}}$
       & $-\frac{\sqrt{3}}{8}$ & $\frac{3}{8}$
       & $-\frac{3}{4\sqrt{2}} $
       & $-1$ & $0$ & $-\frac{1}{\sqrt{3}}$
       & $\frac{3\sqrt{3}}{8}$ & $\frac{3}{8}$
       & $-\frac{3}{4\sqrt{2}} $ \\
$\pi^{-}\Xi^{0}$ & & $1$ & $\sqrt{\frac{2}{3}}$
	   & $-\frac{\sqrt{6}}{8}$ & $-\frac{3}{4\sqrt{2}}$ & $0$
	   & & $-1$ & $-\sqrt{\frac{2}{3}}$
       & $\frac{3\sqrt{6}}{8}$ & $-\frac{3}{4\sqrt{2}}$ & $0$ \\
$\eta\Xi^{-}$ & & & $\frac{1}{3}$
	   & $-\frac{1}{24}$ & $\frac{1}{8\sqrt{3}}$
       & $\frac{1}{4\sqrt{6}}$
       & & & $-\frac{1}{3}$
       & $\frac{1}{8}$ & $\frac{1}{8\sqrt{3}}$
       & $\frac{1}{4\sqrt{6}}$ \\
$K^{-}\Lambda$ & & & & $\frac{5}{6}$ & $\frac{1}{2\sqrt{3}}$
	   & $\frac{1}{\sqrt{6}}$
       & & & & $0$ & $0$ & $0$ \\
$K^{-}\Sigma^{0}$ & & & & & $\frac{1}{2}$ & $0$
       & & & & & $0$ & $\frac{1}{\sqrt{2}}$ \\
$\bar{K}^{0}\Sigma^{-}$  & & & & & & $\frac{1}{2}$ &
       & & & & & $\frac{1}{2}$ \\ \hline
\end{tabular}
    \caption{$B^{d}_{ij}$ and $B^{f}_{ij}(S=-2,Q=-1)$}
    \label{tab:BS-2}
\begin{tabular}{|c|cccccc|cccccc|}\hline
    & \multicolumn{6}{|c|}{$B^{d}_{ij}$}
    & \multicolumn{6}{c|}{$B^{f}_{ij}$}  \\ \hline
    & $\pi^{0}\Xi^{-}$ & $\pi^{-}\Xi^{0}$ & $\eta\Xi^{-}$
    & $K^{-}\Lambda$ & $K^{-}\Sigma^{0}$
    & $\bar{K}^{0}\Sigma^{-}$
    & $\pi^{0}\Xi^{-}$ & $\pi^{-}\Xi^{0}$ & $\eta\Xi^{-}$
    & $K^{-}\Lambda$ & $K^{-}\Sigma^{0}$
    & $\bar{K}^{0}\Sigma^{-}$ \\ \hline
$\pi^{0}\Xi^{-}$ & $0$ & $0$ & $0$
       & $-\frac{1}{8\sqrt{3}}$ & $\frac{1}{8}$
       & $-\frac{1}{4\sqrt{2}} $
       & $0$ & $0$ & $0$
       & $\frac{\sqrt{3}}{8}$ & $\frac{1}{8}$
       & $-\frac{1}{4\sqrt{2}} $ \\
$\pi^{-}\Xi^{0}$ & & $0$ & $0$ & $-\frac{1}{4\sqrt{6}}$
	   & $-\frac{1}{4\sqrt{2}}$ & $0$
	   & & $0$ & $0$ & $\frac{3}{4\sqrt{6}}$
       & $-\frac{1}{4\sqrt{2}}$ & $0$  \\
$\eta\Xi^{-}$ & & & $\frac{4}{3}$ & $\frac{5}{24}$
	   & $-\frac{5}{8\sqrt{3}}$ & $-\frac{5}{4\sqrt{6}}$
       & & & $\frac{4}{3}$ & $-\frac{5}{8}$
       & $-\frac{5}{8\sqrt{3}}$ & $-\frac{5}{4\sqrt{6}}$ \\
$K^{-}\Lambda$  & & & & $\frac{5}{6}$ & $\frac{1}{2\sqrt{3}}$
	   & $\frac{1}{\sqrt{6}}$
       & & & & $0$ & $0$ & $0$\\
$K^{-}\Sigma^{0}$ & & & & & $\frac{1}{2}$ & $0$
	   & & & & & $0$ & $\frac{1}{\sqrt{2}}$
	   \\ 
$\bar{K}^{0}\Sigma^{-}$ & & & & & & $\frac{1}{2}$ &
       & & & & & $\frac{1}{2}$\\ \hline
\end{tabular}
\end{table}%
\begin{table}[tbp]
    \centering
    \caption{$A^{d}_{ij}(S=-1,Q=0)$}
    \label{tab:AdS-1}
\begin{tabular}{|c|cccccccccc|}\hline
     & $K^{-}p$ & $\bar{K}^{0}n$ & $\pi^{0}\Lambda$
    & $\pi^{0}\Sigma^{0}$ &  $\eta\Lambda$
    & $\eta\Sigma^{0}$ & $\pi^{+}\Sigma^{-}$ & $\pi^{-}\Sigma^{+}$
    & $K^{+}\Xi^{-}$ & $K^{0}\Xi^{0}$  \\ \hline
    $K^{-}p$ & $1$ & $\frac{1}{2}$ & $-\frac{\sqrt{3}}{8}$
      & $\frac{3}{8}$ & $-\frac{1}{24}$ & $\frac{1}{8\sqrt{3}}$
      & 0 & $\frac{3}{4}$ & 0 & 0 \\
    $\bar{K}^{0}n$ & & $1$ & $\frac{\sqrt{3}}{8}$ & $\frac{3}{8}$
      & $-\frac{1}{24}$ & $-\frac{1}{8\sqrt{3}}$
      & $\frac{3}{4}$ & 0 & 0 & 0  \\
    $\pi^{0}\Lambda$& & & $\frac{2}{3}$ & 0 & 0 & $\frac{2}{3}$
      & 0 & 0 & $-\frac{\sqrt{3}}{8}$ & $\frac{\sqrt{3}}{8}$ \\
    $\pi^{0}\Sigma^{0}$  & & & & 2 & $\frac{2}{3}$ & 0 & 0 & 0
      & $\frac{3}{8}$ & $\frac{3}{8}$ \\
    $\eta\Lambda$ & & & &  & $\frac{2}{9}$ & 0 & $\frac{2}{3}$
      & $\frac{2}{3}$ & $-\frac{1}{24}$ & $-\frac{1}{24}$  \\
    $\eta\Sigma^{0}$ & & & & & & $\frac{2}{3}$ & 0 & 0
      & $\frac{1}{8\sqrt{3}}$ & $-\frac{1}{8\sqrt{3}}$ \\
    $\pi^{+}\Sigma^{-}$ &  &  &  &  &
      & & $2$ & 0 & $\frac{3}{4}$ & 0 \\
    $\pi^{-}\Sigma^{+}$ &  &  &  &  &
      &  & & $2$& $0$& $\frac{3}{4}$    \\
    $K^{+}\Xi^{-}$ &  &  &  &  &
      &  &  &  & $1$ & $\frac{1}{2}$ \\
    $K^{0}\Xi^{0}$ &  &  &  &  &
      &  &  &  &  & $1$ \\ \hline
\end{tabular}
    \caption{$A^{f}_{ij}(S=-1,Q=0)$}
    \label{tab:AfS-1}
\begin{tabular}{|c|cccccccccc|}\hline
     & $K^{-}p$ & $\bar{K}^{0}n$ & $\pi^{0}\Lambda$
    & $\pi^{0}\Sigma^{0}$ &  $\eta\Lambda$
    & $\eta\Sigma^{0}$ & $\pi^{+}\Sigma^{-}$ & $\pi^{-}\Sigma^{+}$
    & $K^{+}\Xi^{-}$ & $K^{0}\Xi^{0}$  \\ \hline
    $K^{-}p$ & $0$ & $\frac{1}{2}$ & $-\frac{3\sqrt{3}}{8}$
      & $-\frac{3}{8}$ & $-\frac{1}{8}$ & $-\frac{1}{8\sqrt{3}}$
      & 0 & $-\frac{3}{4}$ & 0 & 0 \\
    $\bar{K}^{0}n$ & & $0$ & $\frac{3\sqrt{3}}{8}$ & $-\frac{3}{8}$
      & $-\frac{1}{8}$ & $\frac{1}{8\sqrt{3}}$
      & $-\frac{3}{4}$ & 0 & 0 & 0  \\
    $\pi^{0}\Lambda$& & & $0$ & 0 & 0 & $0$
      & 0 & 0 & $\frac{3\sqrt{3}}{8}$ & $-\frac{3\sqrt{3}}{8}$ \\
    $\pi^{0}\Sigma^{0}$  & & & & 0 & $0$ & 0 & 0 & 0
      & $\frac{3}{8}$ & $\frac{3}{8}$ \\
    $\eta\Lambda$ & & & &  & $0$ & 0 & $0$
      & $0$ & $\frac{1}{8}$ & $\frac{1}{8}$  \\
    $\eta\Sigma^{0}$ & & & & & & 0 & $\frac{2}{\sqrt{3}}$
      & $-\frac{2}{\sqrt{3}}$ & $\frac{1}{8\sqrt{3}}$
      & $-\frac{1}{8\sqrt{3}}$ \\
    $\pi^{+}\Sigma^{-}$ &  &  &  &  &
      & & 0 & 0 & $\frac{3}{4}$ & 0 \\
    $\pi^{-}\Sigma^{+}$ &  &  &  &  &
      &  & & $0$& $0$& $\frac{3}{4}$    \\
    $K^{+}\Xi^{-}$ &  &  &  &  &
      &  &  &  & $0$ & $-\frac{1}{2}$ \\
    $K^{0}\Xi^{0}$ &  &  &  &  &
      &  &  &  &  & $0$ \\ \hline
\end{tabular}
\end{table}
\begin{table}[tbp]
    \centering
    \caption{$B^{d}_{ij}(S=-1,Q=0)$}
    \label{tab:BdS-1}
\begin{tabular}{|c|cccccccccc|}\hline
     & $K^{-}p$ & $\bar{K}^{0}n$ & $\pi^{0}\Lambda$
    & $\pi^{0}\Sigma^{0}$ &  $\eta\Lambda$
    & $\eta\Sigma^{0}$ & $\pi^{+}\Sigma^{-}$ & $\pi^{-}\Sigma^{+}$
    & $K^{+}\Xi^{-}$ & $K^{0}\Xi^{0}$  \\ \hline
    $K^{-}p$ & $1$ & $\frac{1}{2}$ & $-\frac{1}{8\sqrt{3}}$
      & $\frac{1}{8}$ & $\frac{5}{24}$ & $-\frac{5}{8\sqrt{3}}$
      & 0 & $\frac{1}{4}$ & 0 & 0 \\
    $\bar{K}^{0}n$ & & $1$ & $\frac{1}{8\sqrt{3}}$ & $\frac{1}{8}$
      & $\frac{5}{24}$ & $\frac{5}{8\sqrt{3}}$
      & $\frac{1}{4}$ & 0 & 0 & 0  \\
    $\pi^{0}\Lambda$& & & $0$ & 0 & 0 & $0$
      & 0 & 0 & $-\frac{1}{8\sqrt{3}}$ & $\frac{1}{8\sqrt{3}}$ \\
    $\pi^{0}\Sigma^{0}$  & & & & 0 & $0$ & 0 & 0 & 0
      & $\frac{1}{8}$ & $\frac{1}{8}$ \\
    $\eta\Lambda$ & & & &  & $\frac{16}{9}$ & 0 & $0$
      & $0$ & $\frac{5}{24}$ & $\frac{5}{24}$  \\
    $\eta\Sigma^{0}$ & & & & & & $0$ & 0 & 0
      & $-\frac{5}{8\sqrt{3}}$ & $\frac{5}{8\sqrt{3}}$ \\
    $\pi^{+}\Sigma^{-}$ &  &  &  &  &
      & & $0$ & 0 & $\frac{1}{4}$ & 0 \\
    $\pi^{-}\Sigma^{+}$ &  &  &  &  &
      &  & & $0$& $0$& $\frac{1}{4}$    \\
    $K^{+}\Xi^{-}$ &  &  &  &  &
      &  &  &  & $1$ & $\frac{1}{2}$ \\
    $K^{0}\Xi^{0}$ &  &  &  &  &
      &  &  &  &  & $1$ \\ \hline
\end{tabular}
    \caption{$B^{f}_{ij}(S=-1,Q=0)$}
    \label{tab:BfS-1}
\begin{tabular}{|c|cccccccccc|}\hline
     & $K^{-}p$ & $\bar{K}^{0}n$ & $\pi^{0}\Lambda$
    & $\pi^{0}\Sigma^{0}$ &  $\eta\Lambda$
    & $\eta\Sigma^{0}$ & $\pi^{+}\Sigma^{-}$ & $\pi^{-}\Sigma^{+}$
    & $K^{+}\Xi^{-}$ & $K^{0}\Xi^{0}$  \\ \hline
    $K^{-}p$ & $0$ & $\frac{1}{2}$ & $-\frac{\sqrt{3}}{8}$
      & $-\frac{1}{8}$ & $\frac{5}{8}$ & $\frac{5}{8\sqrt{3}}$
      & 0 & $-\frac{1}{4}$ & 0 & 0 \\
    $\bar{K}^{0}n$ & & $0$ & $\frac{\sqrt{3}}{8}$ & $-\frac{1}{8}$
      & $\frac{5}{8}$ & $-\frac{5}{8\sqrt{3}}$
      & $-\frac{1}{4}$ & 0 & 0 & 0  \\
    $\pi^{0}\Lambda$& & & $0$ & 0 & 0 & $0$
      & 0 & 0 & $\frac{\sqrt{3}}{8}$ & $-\frac{\sqrt{3}}{8}$ \\
    $\pi^{0}\Sigma^{0}$  & & & & 0 & $0$ & 0 & 0 & 0
      & $\frac{1}{8}$ & $\frac{1}{8}$ \\
    $\eta\Lambda$ & & & &  & $0$ & 0 & $0$
      & $0$ & $-\frac{5}{8}$ & $-\frac{5}{8}$  \\
    $\eta\Sigma^{0}$ & & & & & & 0 & $0$
      & $0$ & $-\frac{5}{8\sqrt{3}}$ & $\frac{5}{8\sqrt{3}}$ \\
    $\pi^{+}\Sigma^{-}$ &  &  &  &  &
      & & 0 & 0 & $\frac{1}{4}$ & 0 \\
    $\pi^{-}\Sigma^{+}$ &  &  &  &  &
      &  & & $0$& $0$& $\frac{1}{4}$    \\
    $K^{+}\Xi^{-}$ &  &  &  &  &
      &  &  &  & $0$ & $-\frac{1}{2}$ \\
    $K^{0}\Xi^{0}$ &  &  &  &  &
      &  &  &  &  & $0$ \\ \hline
\end{tabular}
\end{table}%


\end{document}